\documentclass{sig-alternate}
\pdfoutput=1
\usepackage{graphicx}
\usepackage[tight,center]{subfigure}
\usepackage{booktabs}
\usepackage{multirow}
\usepackage{amsmath}
\usepackage{amssymb}
\usepackage{amsfonts}
\usepackage{hyperref}
\hypersetup{%
  pdftitle={Strong Regularities in Growth and Decline of Popularity of Social Media Services},
  pdfauthor={Christian Bauckhage and Kristian Kersting},
  pdfsubject={Web-based services, time series analysis},
  pdfkeywords={social media, collective attention, trend prediction},
  colorlinks=true,
  bookmarks=false}
  
\renewcommand{\vec}{\boldsymbol}

\graphicspath{{./}}

\makeatletter
\def\@copyrightspace{\relax}
\makeatother

\begin{document}

\title{Strong Regularities in Growth and Decline \\ of Popularity of Social Media Services}

\numberofauthors{3}

\author{%
\alignauthor
Christian Bauckhage \\
       \affaddr{University of Bonn, \\ Fraunhofer IAIS}\\
       \affaddr{Bonn, Germany}
\alignauthor
Kristian Kersting \\
       \affaddr{TU Dortmund University, \\ Fraunhofer IAIS}\\
       \affaddr{Dortmund, Germany}
}

\maketitle

\begin{abstract}
We analyze general trends and pattern in time series that characterize the dynamics of collective attention to social media services and Web-based businesses. Our study is based on search frequency data available from Google Trends and considers 175 different services. For each service, we collect data from 45 different countries as well as global averages. This way, we obtain more than 8,000 time series which we analyze using diffusion models from the economic sciences. We find that these models accurately characterize the empirical data and our analysis reveals that collective attention to social media grows and subsides in a highly regular and predictable manner. Regularities persist across regions, cultures, and topics and thus hint at general mechanisms that govern the adoption of Web-based services. We discuss several cases in detail to highlight interesting findings. Our methods are of economic interest as they may inform investment decisions and can help assessing at what stage of the general life-cycle a Web service is at.
\end{abstract}




\section{Introduction}

The problem of understanding the dynamics of collective human attention has been called a key scientific challenge for the information age \cite{Wu2007-NAC}. In this paper, we address a specific aspect of this problem and mine search frequency data for common trends and shared characteristics. Our focus is on query logs which summarize the evolution of global and regional interests in \emph{social media services} and we explore to what extend the general dynamics of collective attention apparent from these data can be modeled mathematically.

\begin{figure}[t!]
  \centering
  \subfigure[buzznet]{\includegraphics[width=0.48\columnwidth]{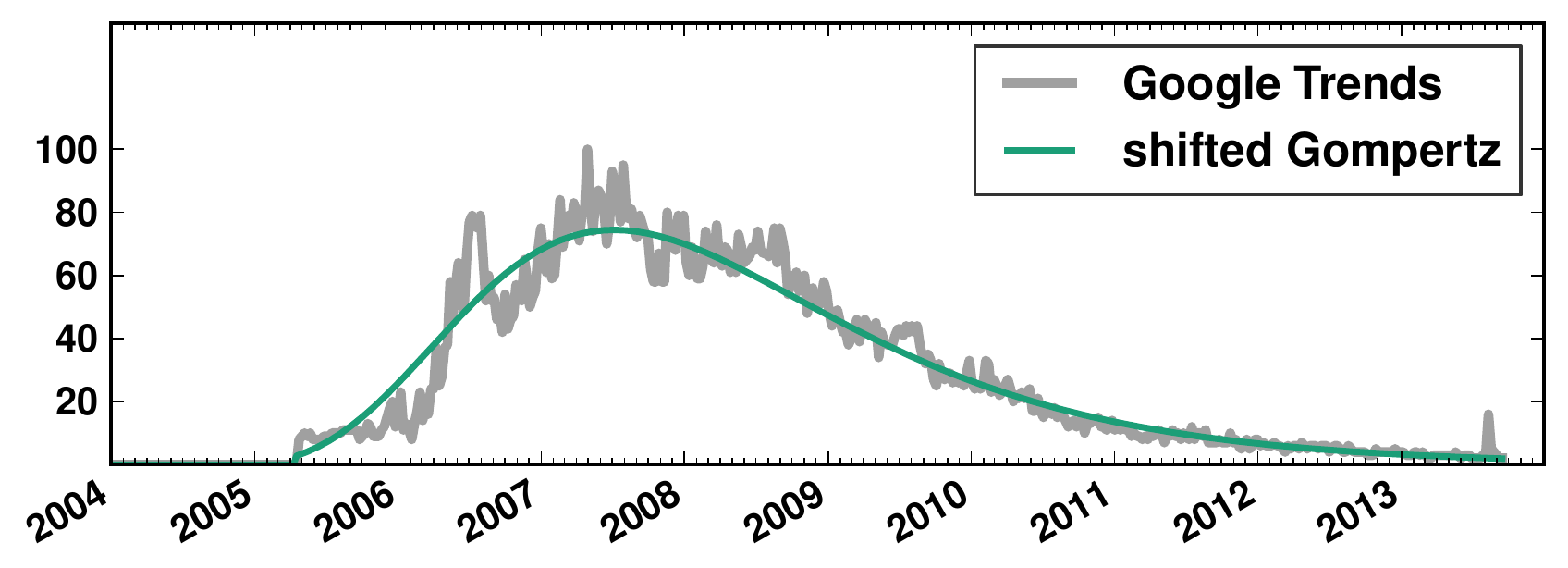}}
  \subfigure[failblog]{\includegraphics[width=0.48\columnwidth]{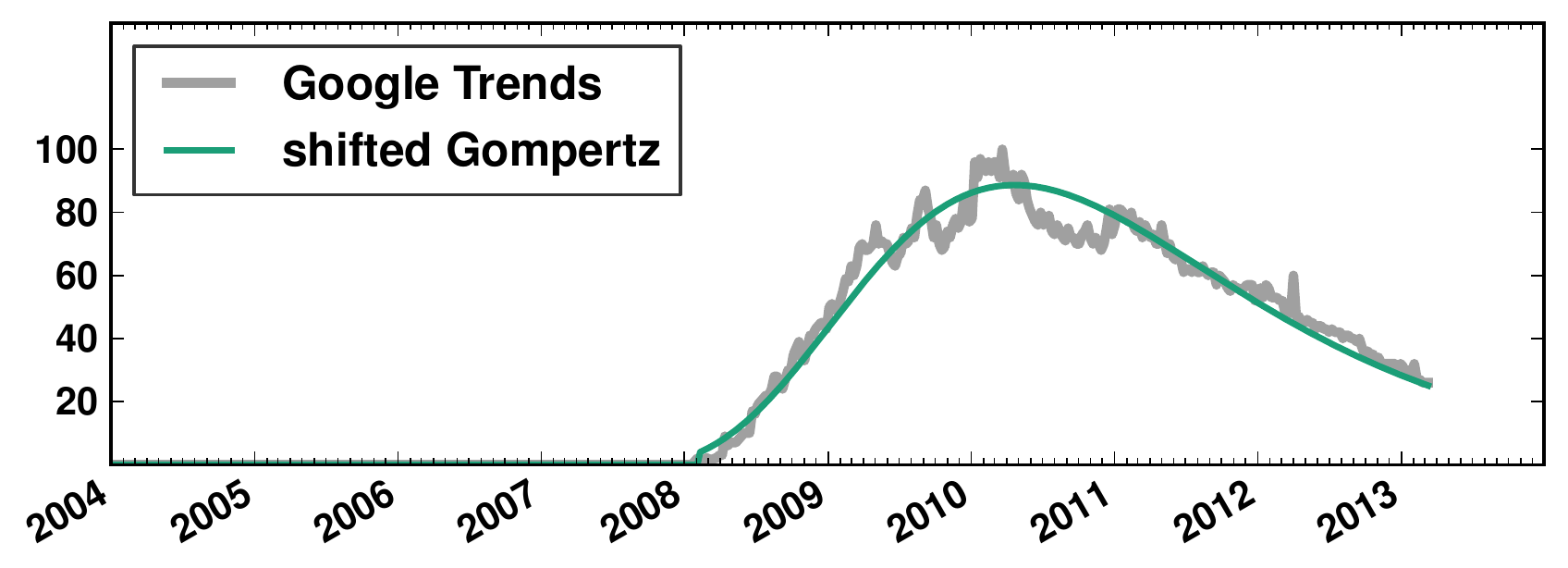}}

  \subfigure[flickr]{\includegraphics[width=0.48\columnwidth]{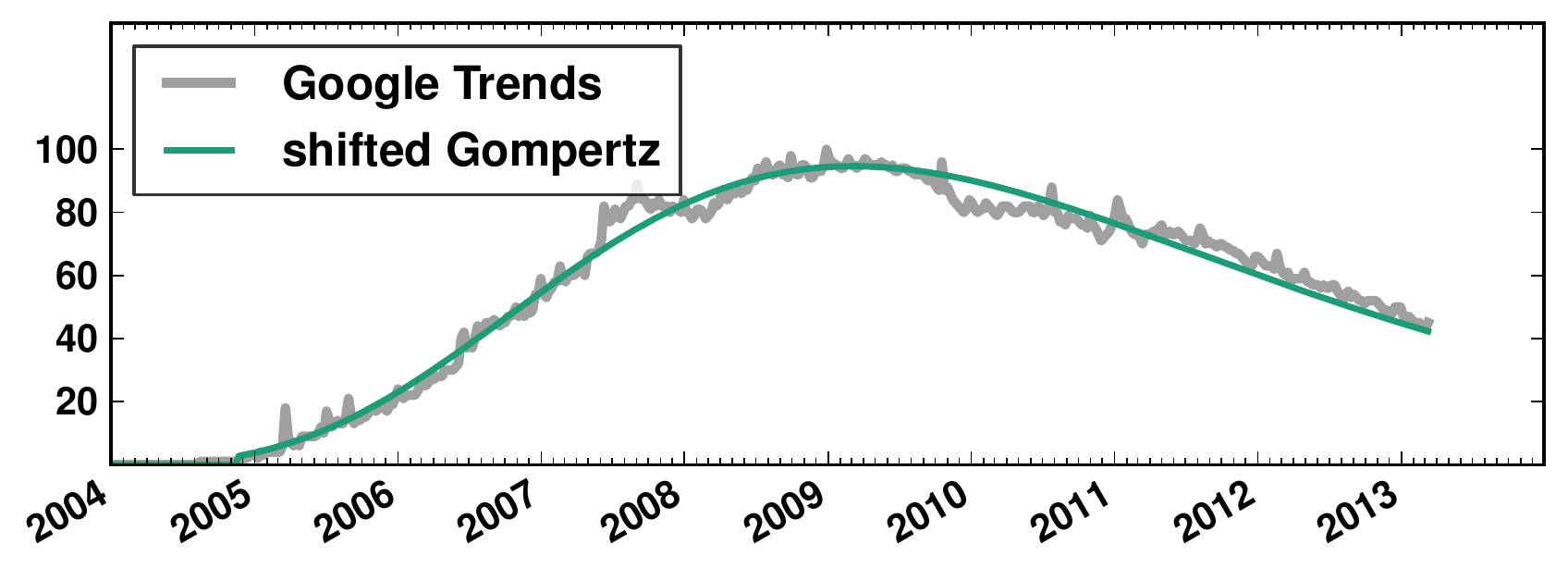}}
  \subfigure[librarything]{\includegraphics[width=0.48\columnwidth]{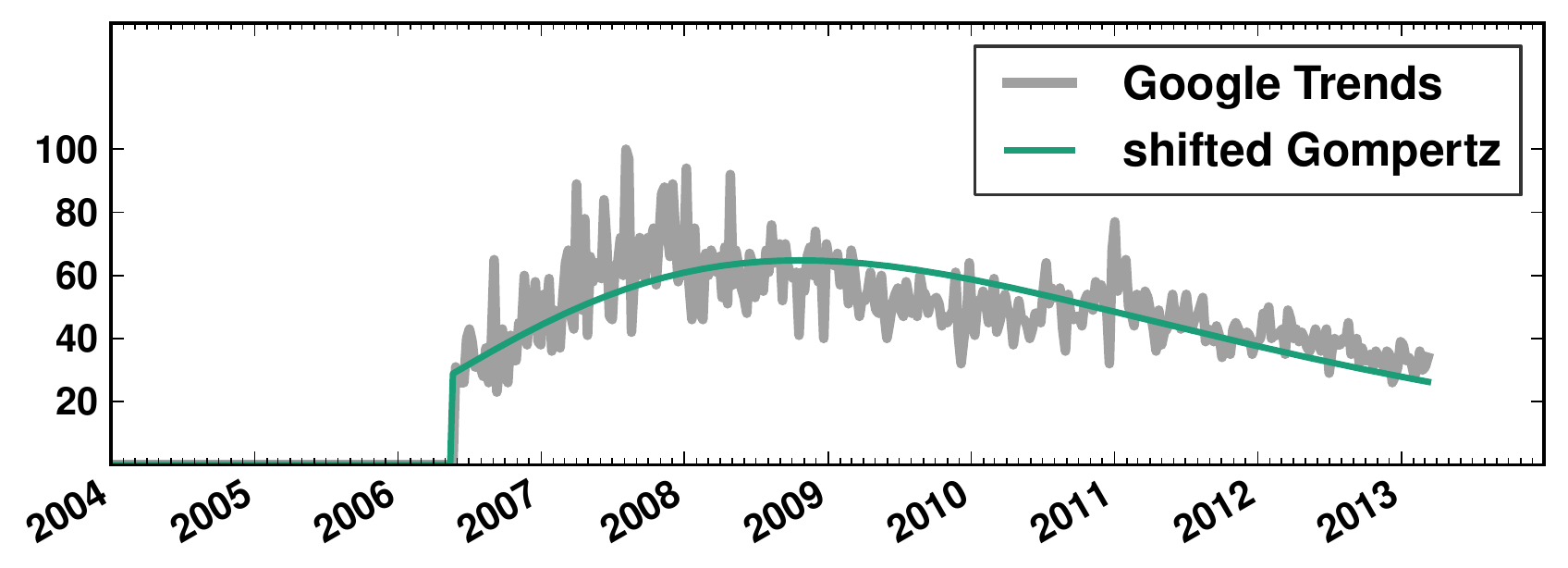}}

  \subfigure[studiVZ]{\includegraphics[width=0.48\columnwidth]{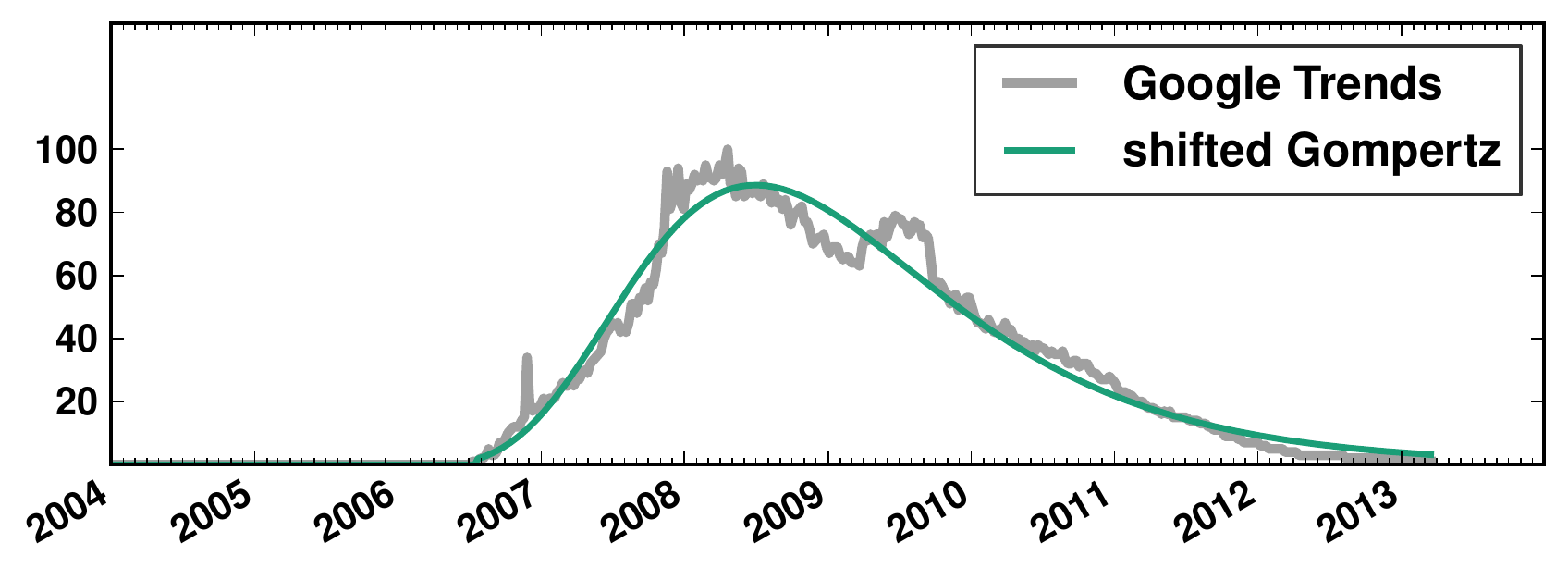}}
  \subfigure[wikipedia]{\includegraphics[width=0.48\columnwidth]{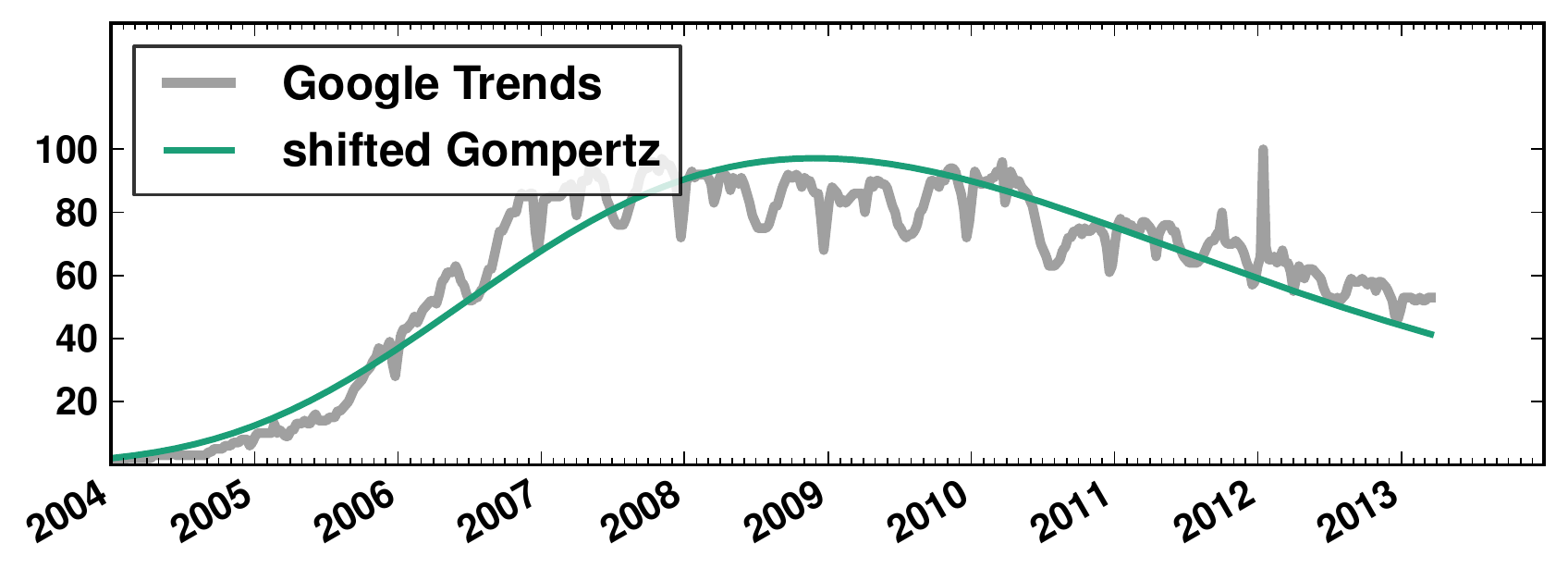}}
  \caption{\label{fig:timeseries} Examples of Google Trends time series which summarize how worldwide searches for different social media services evolve over time. Even though individual curves differ considerably, an appropriately parameterized diffusion model accounts well for the apparent general trends of initial growth and subsequent decline of interest. Results obtained from more than 8,000 temporal signatures of collective attention on the Web indicate that these findings are universal and that interests of large crowds of users follow these patterns regardless of regional, cultural, or linguistics backgrounds.}
\end{figure}

Search frequency analysis is an emerging topic and a growing body of work shows that patterns found in aggregated search data of large populations of Web users can provide  insights into collective concerns, interests, or habits. Results on temporal dynamics of search engine queries are reported from various fields and include data driven models of the spread of diseases \cite{Ginsberg2009-DIE}, accounts of the propagation of news items \cite{Bauckhage2011-III,Bauckhage2013-MMO,Christiansen2012-MTT}, characterizations of the formation of political opinions \cite{Granka2009-ITP}, or predictions of tourism flows \cite{Artola2012-TTF}.

Search frequencies are of particular interest in \emph{nowcasting} which aims at real time monitoring of economic trends and developments \cite{Castle2009-NIN}. Aggregated search behaviors of millions of users yield reliable predictions for sales or general economic indicators \cite{Choi2012-PTP,McLaren2011-UIS}. Temporal changes in search volumes were found to correlate with changes in the behavior of investors \cite{Bordino2013-WSQ,Da2011-ISO} and to allow for predicting abnormal stock returns \cite{Gerow2011-MTW,Joseph2011-FAS}. Accordingly, analysts in the social sciences, public health, or economics are beginning to embrace query log analysis as an alternative to more traditional methods.

The work reported here originates from a project on Web intelligence where we ask for socio-economic motivations for individuals to participate in collective endeavors on the Web. Regarding services, products, and campaigns we investigate approaches that would allow companies or marketeers to recognize whether they need to adjust their strategies in order to remain competitive in the modern Web environment. In particular, we ask to what extent it is possible to predict the future success or adoption of services, products, or marketing messages using collective Web intelligence?

Our paradigm is to mine Web data for possible indicators of trends in collective attention. In this paper, we consider time series obtained from Google Trends which summarize search interests of millions of users worldwide and we focus on temporal signatures that characterize evolving interests in social media. Extending previously published work \cite{Bauckhage2014-CAT}, our contributions are as follows:

1) We briefly review recent results which underline that Google Trends data provide meaningful and reliable proxies for research on how opinions and interests of large crowds and populations evolve over time.

2) We analyze search frequency data from 45 countries related to 175 social media services and Web businesses. Given this comprehensive empirical basis, we perform trend analysis using economic diffusion models and find them to be in excellent agreement with the data. In particular, we find that collective attention to social media as evident from search frequencies evolves according to notably regular patterns. Although microscopic behaviors may be chaotic, general trends apparent in these data typically show simple and highly regular dynamics of growth and decline.

3) We present evidence that this phenomenon persists across regions, cultures, and linguistic backgrounds and we elaborate on several particular examples to highlight several interesting findings. We investigate the potential of our models for forecasting and present qualitative results which indicate that they indeed allow for reasonable predictions of future developments of collective attention.

Next, we discuss the empirical basis of our study. Section~\ref{sec:models} reviews models and methods applied for analysis; results are discussed in section \ref{sec:results}. Section \ref{sec:relatedwork} contrasts our work to the related literature and section~\ref{sec:conclusion} concludes this paper.

\section{Search Frequency Data: A Proxy of Collective Attention}
\label{sec:google}

Our overall goal is to proceed towards a better understanding of the dynamics of collective interests and concerns of large populations of Web users. The empirical basis for the work reported here consists of time series obtained from Google Trends which indicate how search volumes related to specific topics evolve over time.

\subsection{Background}

Google Trends is a publicly accessible service that provides statistics on queries users submitted to Google's search engine. It allows for retrieving weekly summaries of how frequently a query has been used since January 1st 2004. Aggregated statistics are available in form of global averages but can be narrowed down to regional statistics, for instance on the level of individual countries.

Analyzing topic specific search dynamics is an increasingly popular approach in studies on collective preferences
\cite{Artola2012-TTF,Bauckhage2011-III,Bauckhage2013-MMO,Bordino2013-WSQ,Castle2009-NIN,Choi2012-PTP,Da2011-ISO,Gerow2011-MTW,Granka2009-ITP,Joseph2011-FAS,McLaren2011-UIS}
and important questions pertaining to its validity and the significance of search data have been addressed in two recent contributions.

Mellon \cite{Mellon2011-SIA} correlated results from traditional Gallup surveys with Google Trends data and found that, w.r.t.~political and economic issues covered in traditional opinion polls, search frequencies provide accurate proxies of the dynamics of salient public opinions. Teevan et al.~\cite{Teevan2011-UAP} studied how people navigate the Web and found that over 25\% of all queries to search engines are \emph{navigational queries}, i.e.~searches for company names such as \emph{facebook}, \emph{youtube}, or \emph{myspace} that are intended to find and then access particular Web sites. In other words, a large percentage of Web users consistently relies on Google searches rather than on bookmarks or on entering URLs in order to navigate to Web sites. Together these findings thus suggest that data from Google Trends which aggregate information about the search activities of millions of users are indeed indicative of collective interests in Internet services, technical products, or novelties.

\subsection{Data Collection and Preprocessing}

In this paper, we analyze global and regional temporal search statistics related to query terms such as \emph{ebay}, \emph{facebook}, or \emph{youtube} that indicate a populations interest in social media services or Web-based businesses. For potentially ambiguous queries, we retrieve data for different spellings (e.g.~\emph{google plus}, \emph{googleplus}, \emph{google+}, \emph{google +}) and compute their average. In total, we consider data from 45 different countries related to 175 services. As we also retrieve corresponding global search activities, our empirical basis consists of more than 8.000 data sets.

\begin{table}[t!]
\caption{\label{tab:countries} 45 countries considered in this study}
\small
\centering
\begin{tabular}{ll}
\toprule
Africa    & MA, NG, ZA \\
\midrule
Asia      & CN, ID, IN, JP, KR, MY, PH, TH, TW \\
\midrule
Australia & AU, NZ \\
\midrule
Europe    & AT, BE, CH, CZ, DE, DK, ES, FI, FR, GR, \\
          & IE, IL, IT, NL, NO, PL, PT, RU, SE, TR, \\
          & UA, UK \\
\midrule
N-America & CA, MX, US \\
\midrule
S-America & AR, BR, CL, CO, PE, VE \\
\bottomrule
\end{tabular}
\includegraphics[width=0.9\columnwidth]{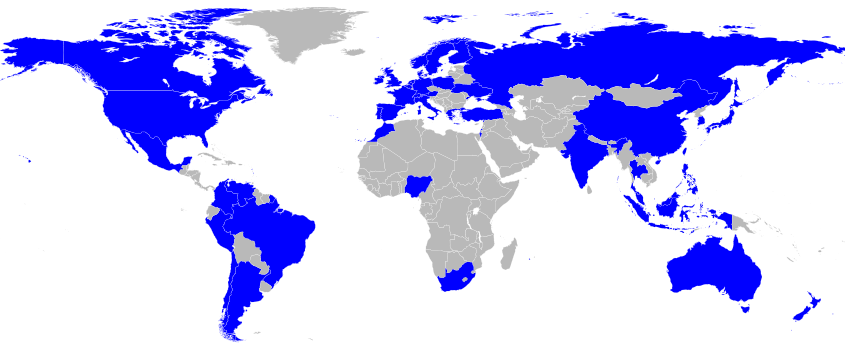}
\end{table}

\begin{table}[t!]
\caption{\label{tab:services} 175 social media services and businesses}
\setlength{\tabcolsep}{5pt}
\small
\centering
\begin{tabular}{llll}
\toprule
43things & flixter & mocospace & studivz \\
adify & fotoki & myheritage & stumbleupon \\
airbnb & fotolog & mylife & svpply \\
aisanavenue & foursquare & myspace & sysomos \\
amazon & friendsreunited & nasza-klasa & taringa \\
amirite & friendster & netlog & techcrunch \\
anobii & gaiaonline & netvibes & technorati \\
asmallworld & getglue & nexopia & tencent-qq \\
badoo & github & odnoklassniki & tripadvisor \\
bebo & gogoyoko & openbc & tripit \\
betfair & goodreads & openid & tuenti \\
bigadda & google+ & orkut & tumblr \\
biip.no & grono & owly & twango \\
bitly & grooveshark & paypal & twitpic \\
blackplanet & groupon & photobucket & twitter \\
bliptv & habbo & pinterest & viadeo \\
boxcryptor & hi5 & plaxo & vimeo \\
busuu & hulu & playdom & virb \\
buzznet & ibibo & posterous & vkontakte \\
cafemom & imgur & qapacity & wakoopa \\
cloob & instagram & quechup & wattpad \\
cotweet & italki & qzone & weeworld \\
cozycot & itsmy & ravelry & weibo \\
craiglist & iwiw & reddit & weread \\
cyworld & janrain & renren & wesabe \\
dailybooth & jiepang & revver & wikia \\
dailymotion & joost & scribd & wikipedia \\
deezer & justin-tv & scvngr & winpalace \\
delicious & kdice & secondlife & wordpress \\
deviantart & kickstarter & seedrs & xanga \\
digg & kiwibox & sevenload & xing \\
disaboom & knitty & shelfari & yelp \\
disqus & lagbook & shopify & youku \\
dontstayin & last.fm & skyblog & youtube \\
dropbox & librarything & skype & zaarly \\
dwolla & linkedin & skyrock & zappos \\
ebay & livejournal & slashdot & zoho \\
elftown & livemocha & slide.com & zoomr \\
elixio & living-social & songza & zooppa \\
epinions & logoworks & sonico & zotero \\
facebook & meebo & soonr & zynga \\
faceparty & meetin & soundcloud &  \\
failblog & mendely & sourceforge &  \\
fetlife & metacafe & spotify &  \\
flickr & mixi & stackoverflow &  \\
\bottomrule
\end{tabular}
\end{table}

The 45 countries considered in this study are listed in Tab.~\ref{tab:countries}. They were selected according to population size, Internet penetration, and availability of  query logs. Note that this sample covers various regions, cultures, and official languages and is deliberately not restricted to countries that are technologically far advanced.

The 175 social media sites and Web businesses we consider are listed in Tab.~\ref{tab:services}. These, too, were chosen according to penetration and profile. Among others, they include general and specialized social networking sites, photo- and video sharing sites, music streaming services, virtual hangouts, (micro-)blogging services, and online retailers, trading platforms, as well as social games providers and thus cover a wide spectrum of social media.

For each combination of country and service, we collect a discrete time series  $\vec{z} = [z_1, z_2, \ldots, z_{483}]$ of weekly search counts $z_t$ from January 2004 to March 2013.

As many services in our sample made their first appearance later than January 2004 (e.g.~\emph{youtube}) and were thus not actively searched for during the whole observation period, we determine individual onset times $t_o$ using CUSUM statistics \cite{Page1954-CIS}. This leaves us with shortened time series $\vec{y} = [y_{t_o}, \ldots, y_{483}]$ which we shift to $y_{t'}$ where $t' = t-t_o$ in order to facilitate statistical analysis.

For query terms related to services that were launched prior to January 2004 (e.g.~\emph{amazon}), we manually determine the number of weeks $T$ between their first public occurrence $t_o$ and January 1st 2004 and consider shifted time series where $t' = t-t_o+T$.

Given these data, we resort to descriptive data mining techniques in order to identify commonalities or significant differences between time series. In particular, we consider three diffusion models which we review in the next section.

\section{Diffusion Models}
\label{sec:models}

Visual inspection of search frequency time series related to social media reveals noticeably common patterns: although, on a microscopic level, collective interest in individual services varies chaotically, macroscopic trends typically show an initial phase of accelerated growth followed by periods of saturation and prolonged decline (see, for instance, the examples in Fig.~\ref{fig:timeseries}).

Skewed temporal distributions like these frequently occur in economics where they indicate buying behaviors or rates of adoption are studied using \emph{diffusion models}. We adhere to this methodology and investigate to what extent simple diffusion models can characterize general trends in our data.

Note that more elaborate approaches such as Gaussian mixtures or kernel techniques might provide more accurate fits. Alas, they typically lack \emph{interpretability} since they yield abstract in terms of (numerous) latent variables without physical meaning. Diffusion models, on the other hand, are deliberately designed to explain time series in terms of intuitive concepts that represent knowledge about everyday life and the real world.

Since we are interested in macroscopic trends, we restrict our analysis to two-parameter models which are unlikely to over-fit the data but will capture its gist. Moreover, they facilitate data exploration and simplify comparisons of sets of time series. In order for this paper to be self contained, this section briefly reviews the three diffusion models we consider.

\subsection{The Bass Model}

In an influential paper, Bass \cite{Bass1969-ANP} proposed a diffusion model to describe how rates of adoption of novel products vary over time. Introducing a parameter $p$ to model a propensity for innovation and a parameter $q$ to model a propensity for imitation, he cast the hazard rate of product adoption as
\begin{equation}
\label{eq:BassHazard}
h(t) = \frac{f(t)}{1 - F(t)} = p + q F(t)
\end{equation}
where $f(t)$ is a probability density and $F(t) = \int_0^t f(\tau) \, d\tau$ is the corresponding cumulative density. Solving the differential equation in (\ref{eq:BassHazard}) leads to the \textbf{Bass distribution}
\begin{equation}
\label{eq:Bass}
f_{\mathcal{BA}}(t \mid p, q) = \frac{(p+q)^2}{p} \frac{e^{-(p+q) t}}{\left( 1 + \frac{q}{p} e^{-(p+q) t}\right)^2}.
\end{equation}

Depending on the choice of $p$ and $q$, this distribution can assume a variety of shapes. In particular, for $q > p$, it will increase to a maximum before decreasing to zero. This becomes explicit by writing (\ref{eq:BassHazard}) as $f(t) = p + q F(t) - q F^2(t)$ which exposes the adoption rate to result from composing two antagonistic processes: a propensity $p + q F(t)$ to grow countered by a propensity $q F^2(t)$ to decline.

We include the Bass model in our analysis because it often accurately models sales and thus may also be able explain collective attention dynamics on the Web.

\subsection{The Shifted Gompertz Model}

As our second model, we consider the \textbf{shifted Gompertz distribution}
\begin{equation}
\label{eq:SGompertz}
f_{\mathcal{SG}}(t \mid \beta, \eta) = \beta e^{-\beta t} e^{-\eta e^{-\beta t}} \Bigl( 1 + \eta \bigl(1 - e^{-\beta t}\bigr) \Bigr)
\end{equation}
where $t, \beta, \eta \geq 0$. It was introduced by Bemmaor \cite{Bemmaor1994-MTD} who showed that the Bass model results from compounding the shifted Gompertz with an Exponential distribution, i.e.
\begin{equation}
\label{eq:BassMixture}
f_{\mathcal{BA}}(t \mid p, q) = \int_0^\infty f_{\mathcal{SG}}(t \mid \beta, \eta) \, \frac{e^{-\frac{\eta}{\sigma}}} {\sigma} \, d\eta
\end{equation}
such that $p = \beta / (1+\sigma)$ and $q = p \sigma$. This reveals a latent coupling of the Bass parameters $p$ and $q$ due to taking the average over the shape parameter $\eta$ of the shifted Gompertz. Bemmaor's shifted Gompertz therefore provides a more flexible characterization of adoption dynamics and we explore its merits in our experiments below.

\subsection{The Weibull Model}

The \textbf{Weibull distribution} is the type III extreme value distribution and often applied as a life-time model \cite{Rinne2008-TWD}. Its probability density function is defined for $t \in [0,\infty)$ and given by
    \begin{equation}
    \label{eq:Weibull}
    f_\mathcal{WB}(t \mid \kappa, \lambda) =
    \frac{\kappa}{\lambda} \Bigl ( \frac{t}{\lambda} \Bigr)^{\kappa-1} e^{-(t/\lambda)^\kappa}
    \end{equation}
where $\kappa$ and $\lambda$ determine shape and scale. For $\kappa = 1$, the Weibull coincides with the Exponential and, for $\kappa \approx 3.5$, it approaches the Standard Normal.

Studying the dynamics of Internet memes, Bauckhage et al.~\cite{Bauckhage2013-MMO} pointed out that the Weibull, too, implicitly couples two antagonistic growth dynamics. This can be seen from considering its cumulative density function
\begin{equation}
\label{eq:WeibullCDF}
F_\mathcal{WB}(t \mid \kappa, \lambda) = 1 - e^{-(t/\lambda)^\kappa}.
\end{equation}
Setting  $\alpha = (\tfrac{1}{\lambda})^\kappa$ for brevity, rearranging the terms in (\ref{eq:WeibullCDF}), and substituting into (\ref{eq:Weibull}) yields $f(t) = \alpha \kappa t^{\kappa-1} - \alpha \kappa t^{\kappa-1} F(t)$. Considered as a diffusion model, the Weibull distribution thus combines a propensity $\alpha \kappa t^{\kappa-1}$ for collective attention to a service or product to grow with a propensity $\alpha \kappa t^{\kappa-1} F(t)$ for attention to subside. In passing, we note that by letting $\alpha = \alpha(t) = q F(t)$ and setting $\kappa = 1$, the Weibull and the Bass model are related as $f_{\mathcal{BA}}(t) - p = f_{\mathcal{WB}}(t)$.

\subsection{Model Fitting}

When applying the above diffusion models to analyze temporal signatures of collective attention on the Web, we must cope with the fact that neither model provides a closed form solution for the maximum likelihood estimates of their parameters. Addressing this issue and aiming at high efficiency for large scale processing, we propose the use of \emph{multinomial maximum likelihood} techniques.

Throughout, we fit continuous distributions $f(t \mid \theta_1, \theta_2)$ to discrete series of frequency counts $y_1, \ldots, y_m$ grouped into $m$ distinct intervals $(t_0,t_1]$, $(t_1,t_2]$, $\ldots$, $(t_{m-1}, t_m]$. To devise an efficient algorithm for estimating optimal model parameters $\theta_1^*$ and $\theta_2^*$, we note that a histogram $h(y_1, \ldots, y_m)$ of counts can be thought of as a multinomial distribution
\begin{equation}
h(y_1, \ldots, y_m) = n!\prod_i \frac{p_i^{y_i}}{y_i!}
\end{equation}
where $n = \sum_i y_i$. Since the cumulative density of the model distribution is
\begin{equation}
F(t) = F(t \mid \theta_1, \theta_2) = \int_0^t f(\tau \mid \theta_1, \theta_2) \; d\tau,
\end{equation}
the probabilities $p_i$ of the multinomial can be expressed as $p_i(\theta_1, \theta_2) = F(t_i) - F(t_{i-1})$ so that $\sum_i p_i = F(t_m) - F(t_0)$. Accordingly,
the likelihood for a discrete, truncated time series $y_1, \ldots, y_m$ is given by
\begin{equation*}
L(\theta_1, \theta_2) = \frac{n!}{F(t_m) - F(t_0)} \prod_i \frac{p_i(\theta_1, \theta_2) ^{y_i}}{y_i!}
\end{equation*}
and maximum likelihood estimates of $\theta_1$ and $\theta_2$ result from computing the roots of $\nabla_{\vec{\theta}} \log L$. Again, this may not lead to closed form solutions but may require numerical optimization. To this end, we apply an efficient, iterative weighted least squares scheme
\begin{equation*}
\sum_i w_i \bigl ( y_i - n p_i (\theta_1, \theta_2) \bigr )^2.
\end{equation*}
which regresses the $y_i$ onto their expectations $n p_i$ and requires to update the weights $w_i = ( n p_i )^{-1}$ in each iteration.

In addition to computational convenience, this approach is robust and has the property that, for $p_i = p_i (\theta_1^*, \theta_2^*)$, the final residual sum of squares follows a $\chi^2$ statistic \cite{Jennrich1975-MLE}. We thus resort to the $\chi^2$-test for goodness of fit (GoF) testing. Yet, we note that the $\chi^2$-test may underestimate the quality of fits to time series \cite{Gleser1983-TEO} so that the results reported below may improve even further if more elaborate tests were used.

\section{Empirical Results}
\label{sec:results}

This section presents and discusses trend analysis results for our data set of about 8,000 social media related search frequency signatures. In order to illustrate several arguably important findings, we compare results obtained for distinct countries, regions of the world, linguistic backgrounds, and types of service in form of small case studies.

\subsection{Time to Adoption}

In a preparatory analysis, we gather statistics as to times to adoption of social media in different countries. For each service in our data set, we determine its global onset, i.e.~the point in time at which it first became visible in Google's search frequency data. Then, for every country in our data, we determine the delay $\Delta t$ (in days) between the service's global onset and its onset in the country. Finally, we compute the mean ($\mu$) and median ($m$) delay per country in order to perform comparisons.

Table~\ref{tab:adoption} ranks the 45 countries considered according to their mean- and median times to adoption; the world map at the bottom of the table shows a heat map visualization of median times to adoption. Together with Japan, countries from the western world lead both rankings. With respect to both metrics, the US is the country where social media most quickly achieve noticeable rates of adoption. This is less surprising since many popular social media services such as \emph{facebook} are based in the US and thus may gather an American audience faster than a global one.

\begin{table}[t]
\caption{\label{tab:adoption} Rankings of countries w.r.t.~mean ($\mu$) and median ($m$) time to adoption of a novel service}
\centering
\begin{tabular}{rcc|rcc|rcc}
\toprule
& $\mu$ & $m$ & & $\mu$ & $m$ & & $\mu$ & $m$ \\
\midrule
1. & US & US & 16. & NL & IE & 31. & NZ & AR \\
2. & UK & UK & 17. & AT & MX & 32. & TW & DK \\
3. & FR & CA & 18. & IE & AT & 33. & DK & NO \\
4. & CA & FR & 19. & MX & PL & 34. & ZA & ZA \\
5. & JP & DE & 20. & PL & MY & 35. & CN & TW \\
6. & DE & JP & 21. & PH & IL & 36. & CO & CO \\
7. & AU & ES & 22. & IN & NZ & 37. & TH & VE \\
8. & IT & IT & 23. & PT & BR & 38. & NO & GR \\
9. & ES & NL & 24. & IL & PH & 39. & GR & NG \\
10. & BE & BE & 25. & MA & CL & 40. & CZ & CZ \\
11. & SE & AU & 26. & VE & MA & 41. & ID & UA \\
12. & MY & SE & 27. & CL & PE & 42. & UA & ID \\
13. & PE & FI & 28. & AR & IN & 43. & KR & TH \\
14. & FI & PT & 29. & TR & TR & 44. & RU & RU \\
15. & CH & CH & 30. & BR & CN & 45. & NG & KR \\
\bottomrule
\end{tabular}
\includegraphics[width=\columnwidth]{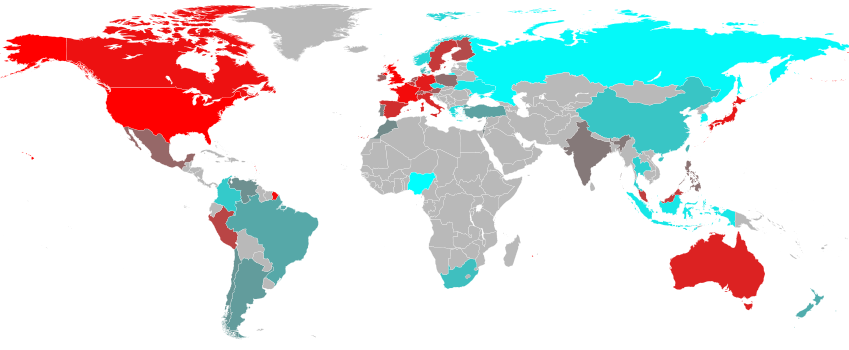}
\end{table}

\begin{figure}[t]
\includegraphics[width=1.\columnwidth]{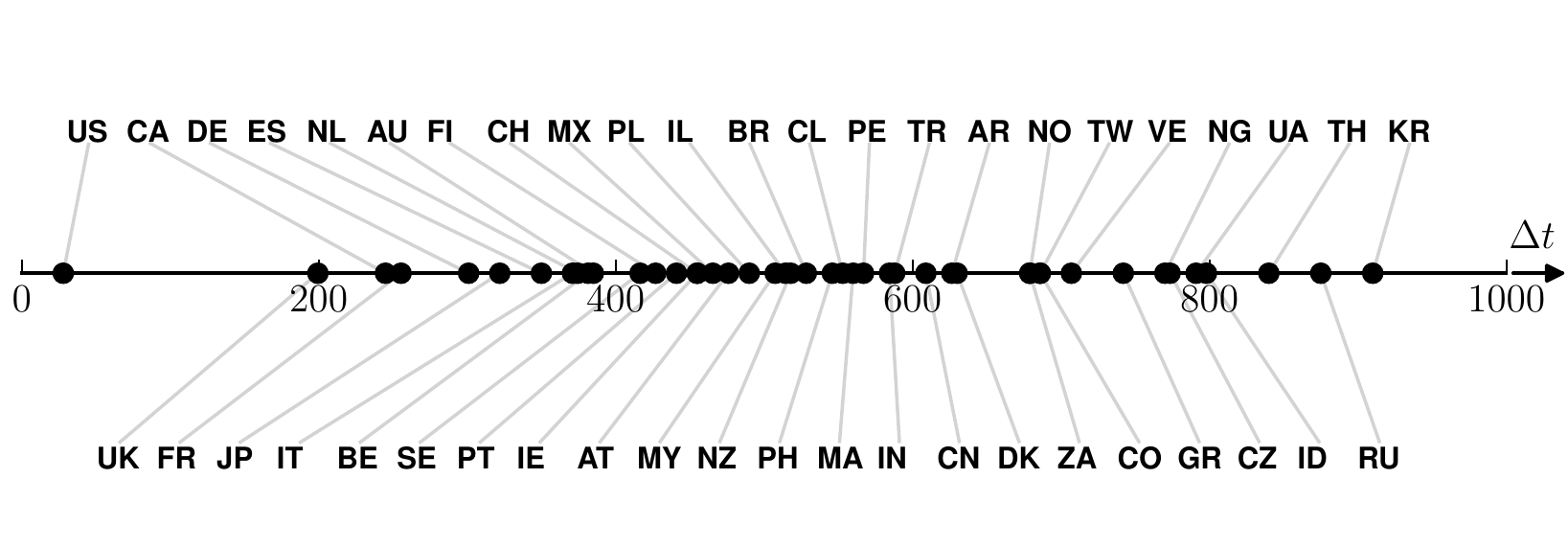}
\caption{\label{fig:timeline} Time line showing median times to adoption (in days) of social media for different countries.}
\end{figure}

A less anticipated finding comes from looking at Fig.~\ref{fig:timeline} which plots median times to adoption along the time axis. The delay between the US and the next fastest adopting country, the UK, amounts to more than 200 days. For the majority of countries in our study, we find that Web-based social media achieve noticeable rates of adaption between 400 and 600 days after their launch or first observable onset. At first sight, it thus appears surprising to find South Korea, a technologically highly advanced nation, to lag behind in this statistic. Yet, this can be attributed to peculiar aspects of South Korean Web culture which features many social media such as \emph{cyworld} or \emph{me2day} that are very popular within the country but rather unknown elsewhere.

Findings like these further underline that search frequency signatures indeed provide plausible proxies for the study of collective attention on the Web. Next, we therefore address aspects of attention dynamics expressed in query log data.

\subsection{Attention Dynamics}

\begin{table}[t!]
\caption{\label{tab:gofreg} Goodness of fit w.r.t.~regions of the world}
\setlength{\tabcolsep}{4pt}
\small
\centering
\begin{tabular}{lcccccc}
\toprule
\multirow{2}{*}{region} & \multicolumn{2}{c}{$f_\mathcal{SG}$} & \multicolumn{2}{c}{$f_\mathcal{BA}$} & \multicolumn{2}{c}{$f_\mathcal{WB}$} \\
\cmidrule(rl){2-3} \cmidrule(rl){4-5} \cmidrule(rl){6-7}
& $\langle p \rangle$ & $p > 0.05$ & $\langle p \rangle$ & $p > 0.05$ & $\langle p \rangle$ & $p > 0.05$\\
\midrule
Africa     & 0.61 & 68\% & 0.55 & 62\% & 0.50 & 57\% \\
Asia       & 0.57 & 63\% & 0.49 & 54\% & 0.48 & 53\% \\
Australia  & 0.66 & 70\% & 0.53 & 59\% & 0.50 & 58\% \\
Europe     & 0.59 & 65\% & 0.48 & 51\% & 0.56 & 54\% \\
N-America  & 0.54 & 57\% & 0.44 & 50\% & 0.39 & 44\% \\
S-America  & 0.65 & 71\% & 0.54 & 59\% & 0.55 & 62\% \\
 \midrule
worldwide  & 0.59 & 64\% & 0.50 & 55\% & 0.47 & 53\% \\
\bottomrule
\end{tabular}
\end{table}

\begin{table}[t!]
\caption{\label{tab:goflang} Goodness of fit w.r.t.~languages of the world}
\setlength{\tabcolsep}{4pt}
\small
\centering
\begin{tabular}{lcccccc}
\toprule
\multirow{2}{*}{language} & \multicolumn{2}{c}{$f_\mathcal{SG}$} & \multicolumn{2}{c}{$f_\mathcal{BA}$} & \multicolumn{2}{c}{$f_\mathcal{WB}$} \\
\cmidrule(rl){2-3} \cmidrule(rl){4-5} \cmidrule(rl){6-7}
& $\langle p \rangle$ & $p > 0.05$ & $\langle p \rangle$ & $p > 0.05$ & $\langle p \rangle$ & $p > 0.05$\\
\midrule
English    & 0.55 & 58\% & 0.44 & 49\% & 0.39 & 45\% \\
Spanish    & 0.63 & 68\% & 0.52 & 56\% & 0.54 & 60\% \\
Portuguese & 0.60 & 67\% & 0.50 & 56\% & 0.47 & 51\% \\
Russian    & 0.68 & 76\% & 0.58 & 66\% & 0.69 & 76\% \\
French     & 0.55 & 60\% & 0.46 & 51\% & 0.39 & 45\% \\
German     & 0.58 & 64\% & 0.47 & 52\% & 0.47 & 54\% \\
Chinese    & 0.50 & 52\% & 0.42 & 46\% & 0.43 & 47\% \\
Japanese   & 0.42 & 52\% & 0.38 & 44\% & 0.31 & 38\% \\
Hindi      & 0.57 & 64\% & 0.47 & 54\% & 0.48 & 52\% \\
\midrule
average    & 0.57 & 62\% & 0.47 & 52\% & 0.45 & 51\% \\
\bottomrule
\end{tabular}
\end{table}

In our main analysis, we apply the economic diffusion models from section~\ref{sec:models} in order to mine our data for shared characteristics or noteworthy exceptions.

Table~\ref{tab:gofreg} presents Goodness-of-Fit (GoF) results for all three models in terms of $p$-value statistics obtained from $\chi^2$-tests.  To produce these statistics, data from different countries were grouped into clusters representing continents and the models were evaluated for each cluster. For the shifted Gompertz, average $p$-values (the higher the better) significantly exceed $0.5$. This holds for fits to data which reflect worldwide interests as well as for fits to continent specific data. Moreover, at a significance level of $5\%$, we find the shifted Gompertz to provide accurate fits for the majority of our data. In terms of overall GoF, the Bass and the Weibull perform slightly worse, yet both models yield statistically significant fits for most of the data, too.

Table~\ref{tab:goflang} provides an alternative view on our data. While Tab.~\ref{tab:gofreg} shows results w.r.t.~geographic regions, Tab.~\ref{tab:goflang} lists GoF results w.r.t.~major languages spoken across the world. Data from different countries were grouped into clusters representing official languages and the three diffusion models were evaluated for each cluster. Apparently, the results in Tab.~\ref{tab:goflang} mimic those in Tab.~\ref{tab:gofreg}. Quality and significance of fits are comparable and the shifted Gompertz again provides the most accurate explanation.

These results are interesting and important for they suggest that \emph{the dynamics of collective attention apparent from search frequency data can be accurately described in terms of diffusion models}. Moreover, they indicate that, around the world, \emph{collective attention to social media evolves similarly and independent of regions of origin or cultural backgrounds of crowds of Web users}.

\begin{figure*}[t]
\centering
\subfigure[amazon]{\includegraphics[width=0.49\columnwidth]{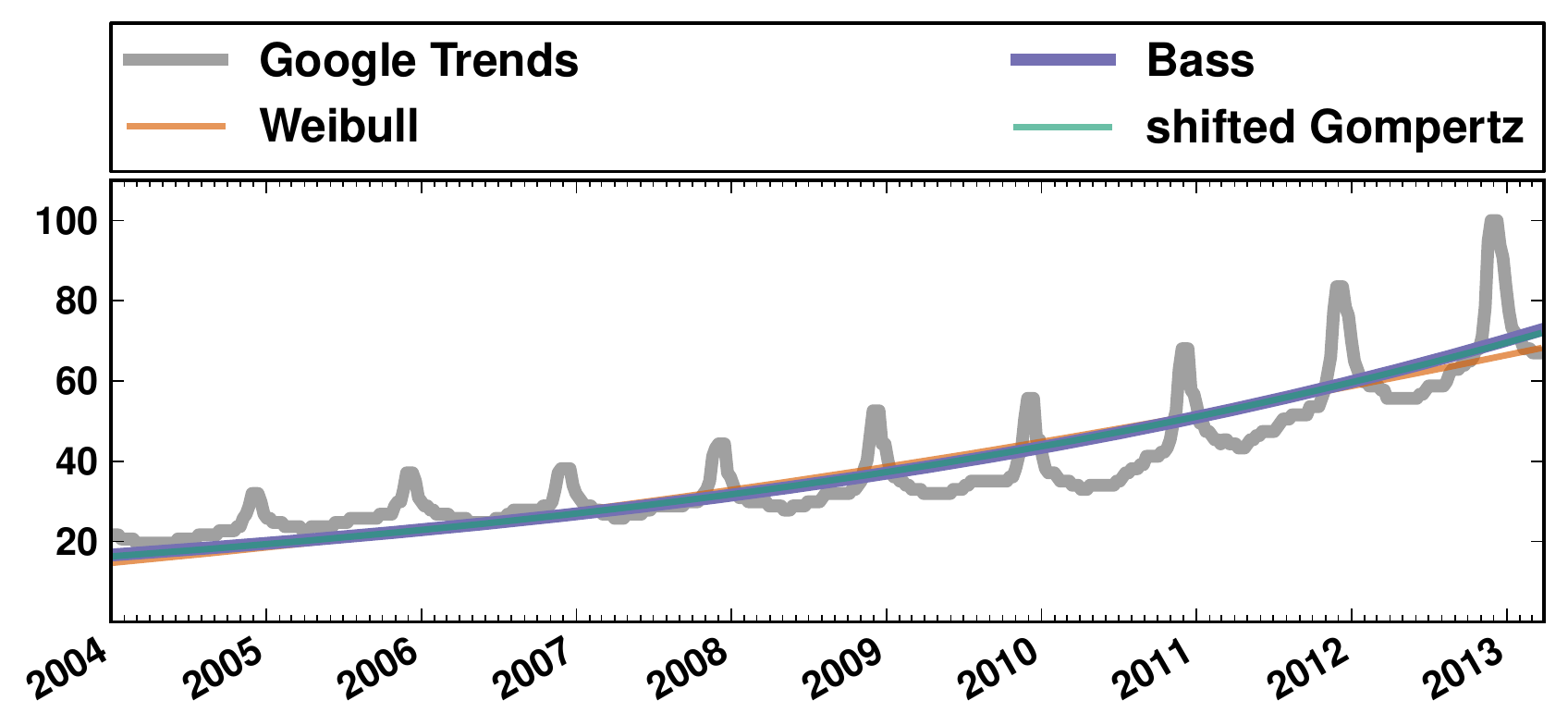}}
\subfigure[craiglist]{\includegraphics[width=0.49\columnwidth]{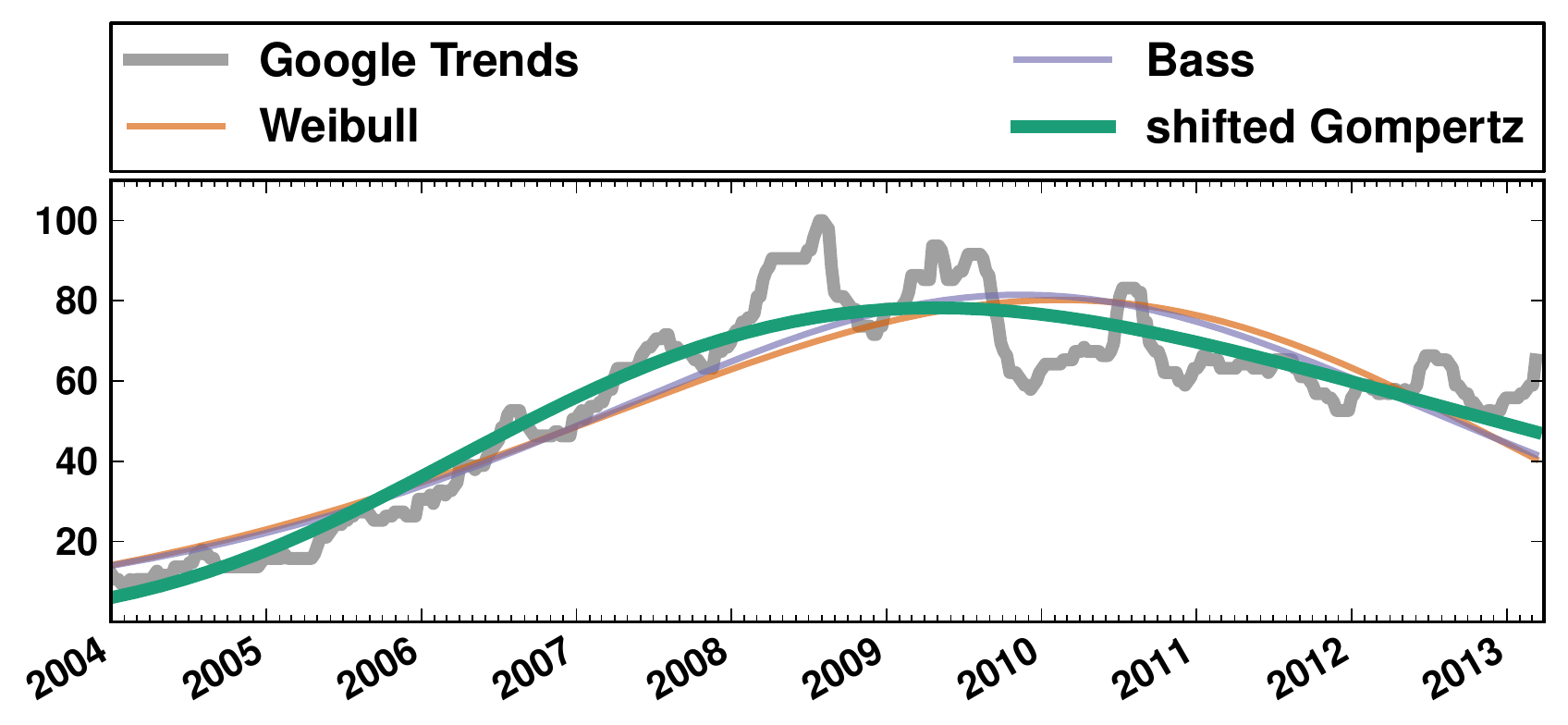}}
\subfigure[ebay]{\includegraphics[width=0.49\columnwidth]{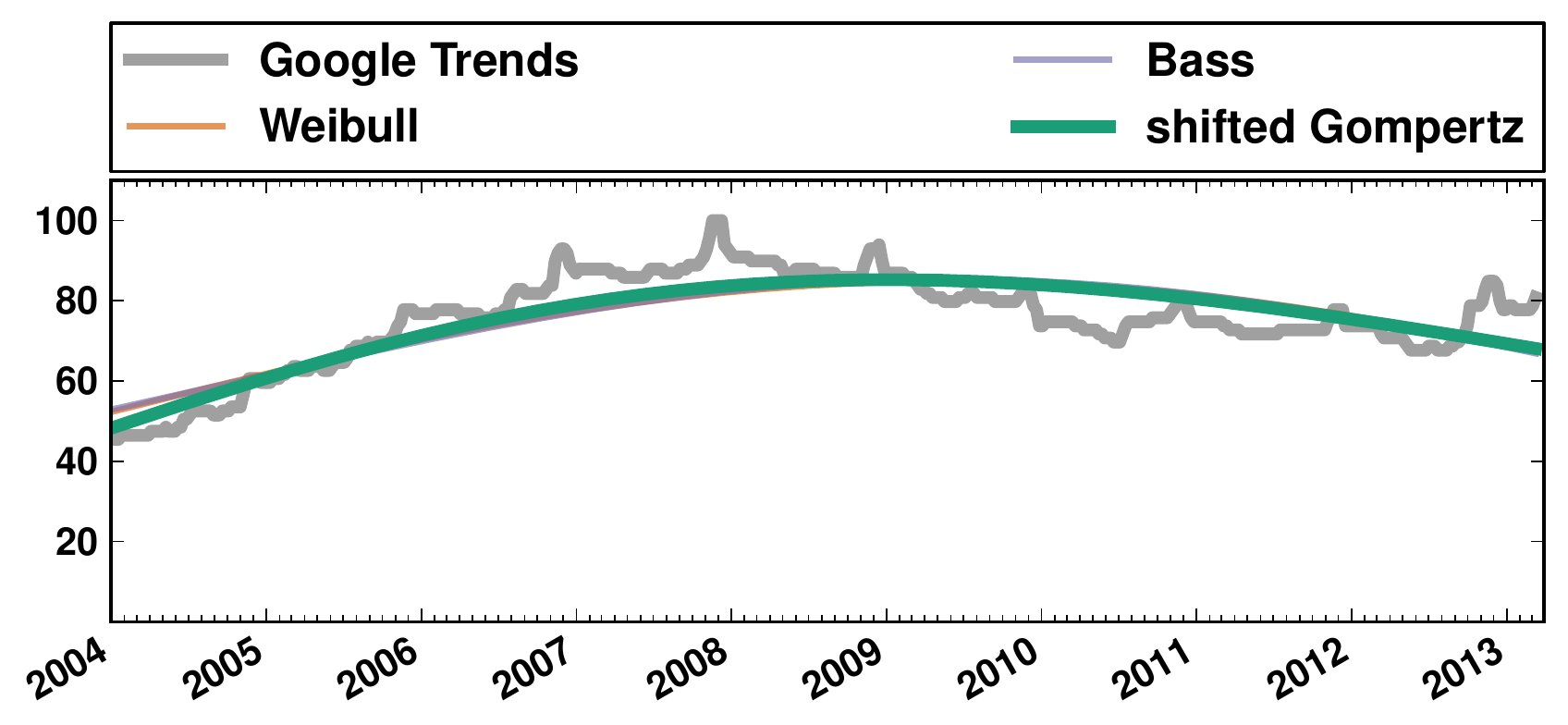}}
\subfigure[facebook]{\includegraphics[width=0.49\columnwidth]{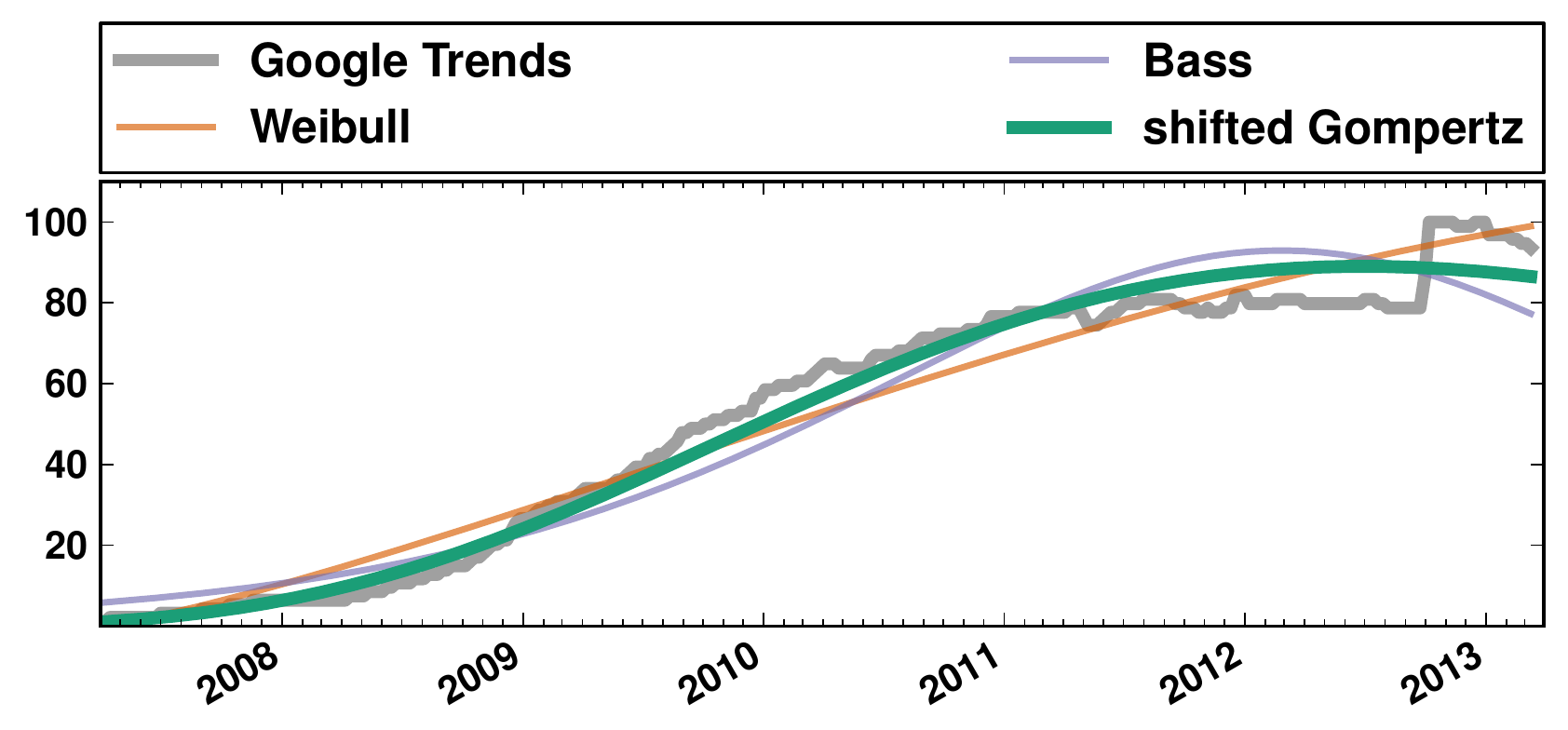}}

\subfigure[google+]{\includegraphics[width=0.49\columnwidth]{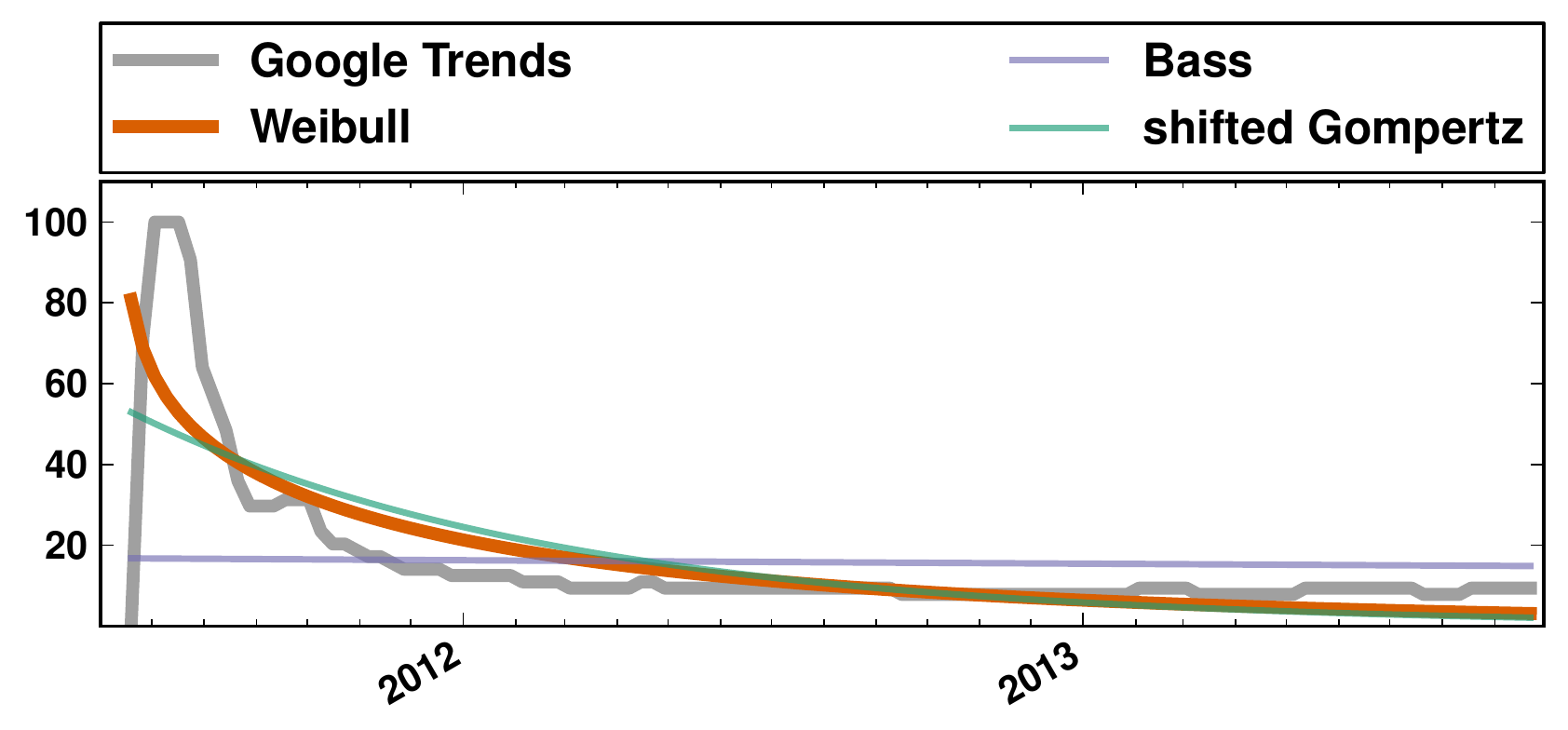}}
\subfigure[myspace]{\includegraphics[width=0.49\columnwidth]{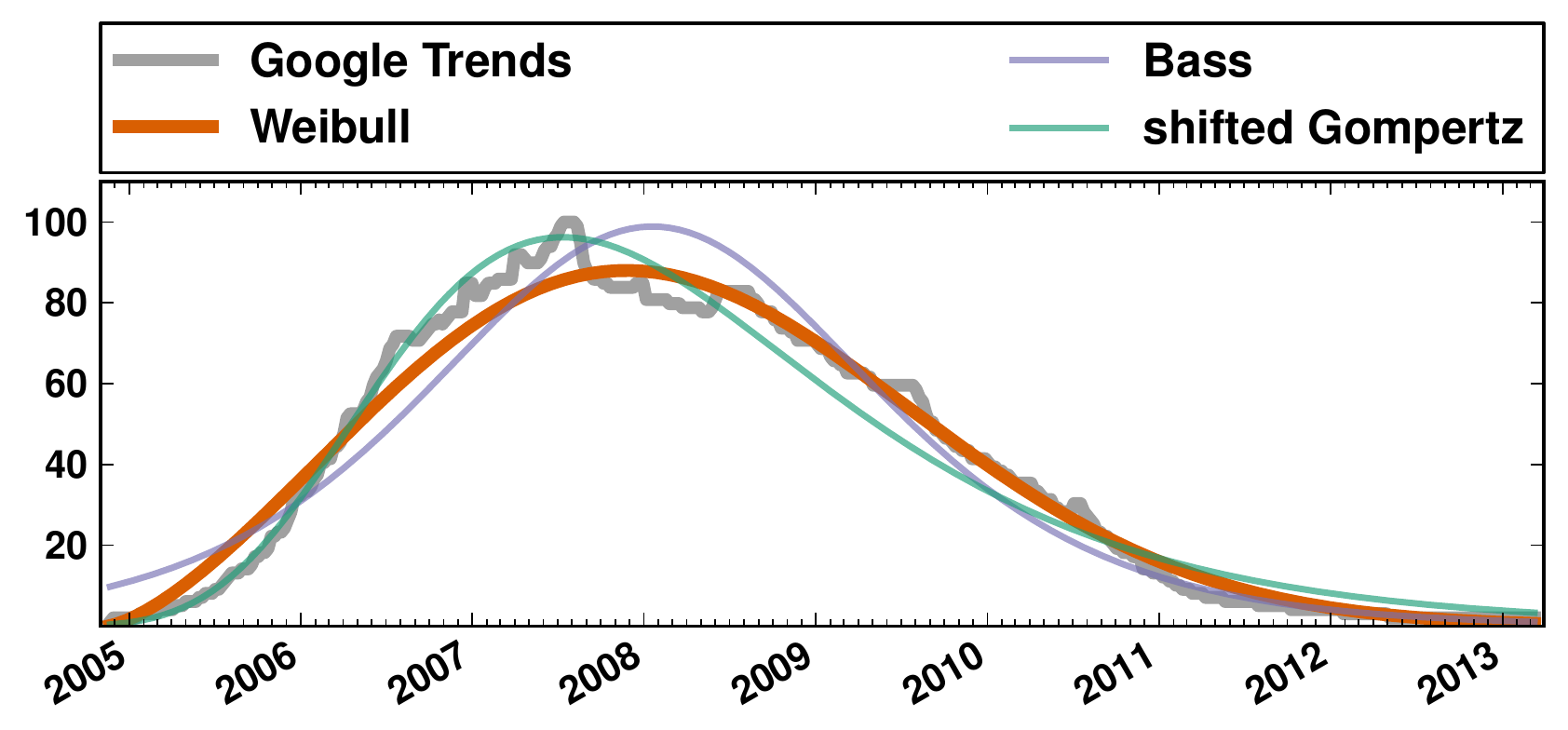}}
\subfigure[youtube]{\includegraphics[width=0.49\columnwidth]{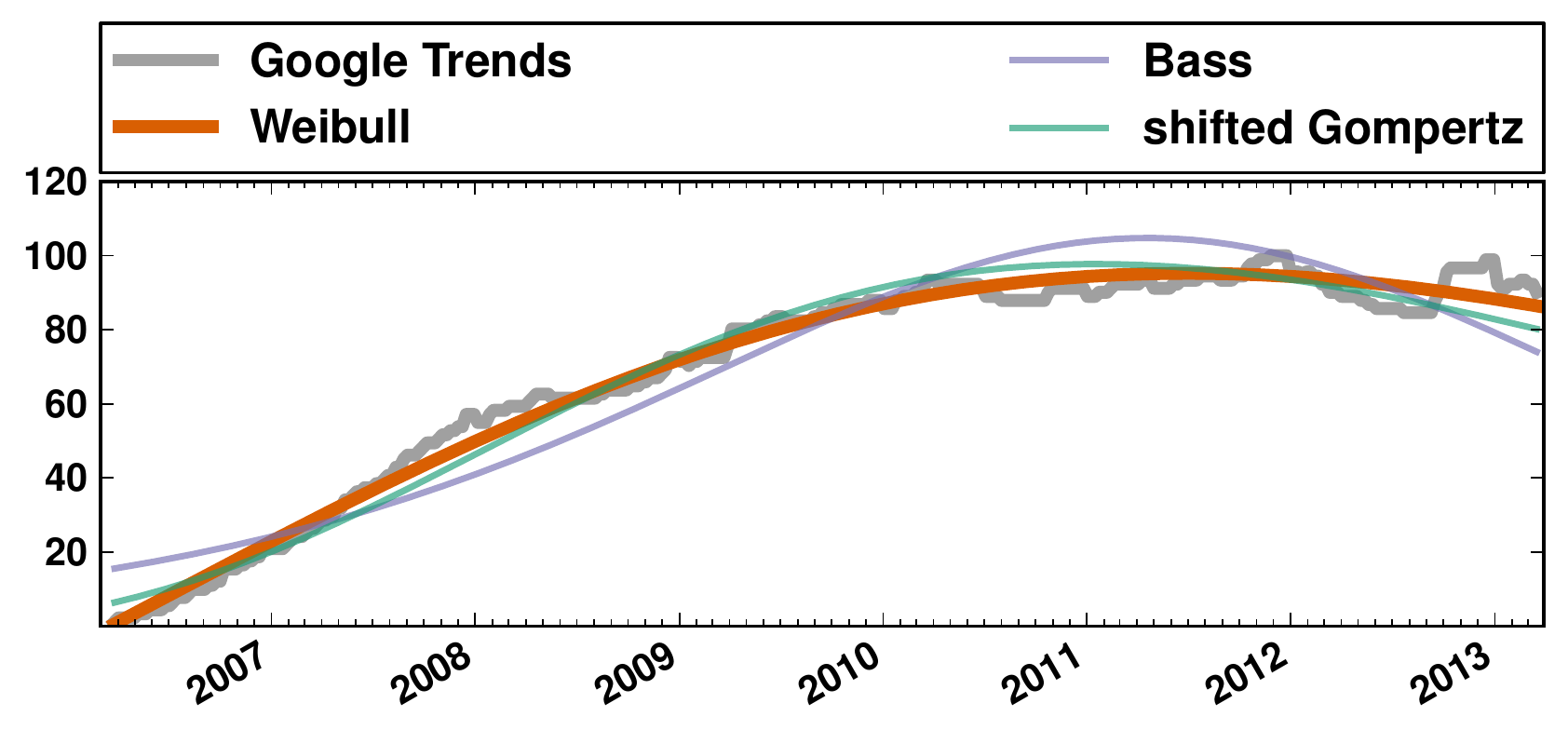}}
\subfigure[twitter]{\includegraphics[width=0.49\columnwidth]{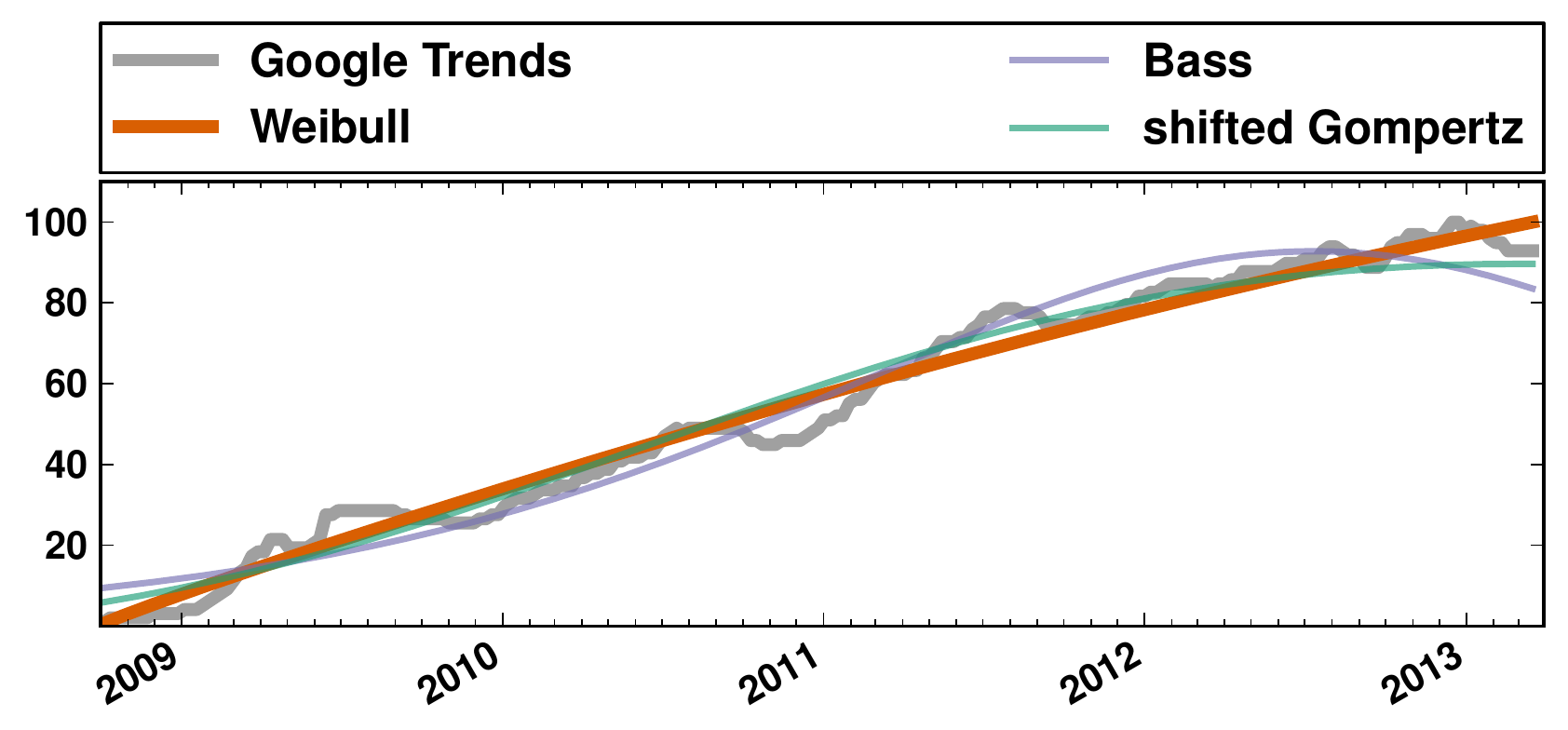}}
\caption{\label{fig:fitcmp} Exemplary visualizations of how the three diffusion models  (Bass, shifted Gompertz, and Weibull) fit general trends in temporal signatures of worldwide query logs related to several popular and well known social media services and Web-based businesses; the respective best fitting model is emphasized.}
\end{figure*}

\begin{figure*}[t]
\centering
\subfigure[Bass model]{\includegraphics[width=.32\textwidth]{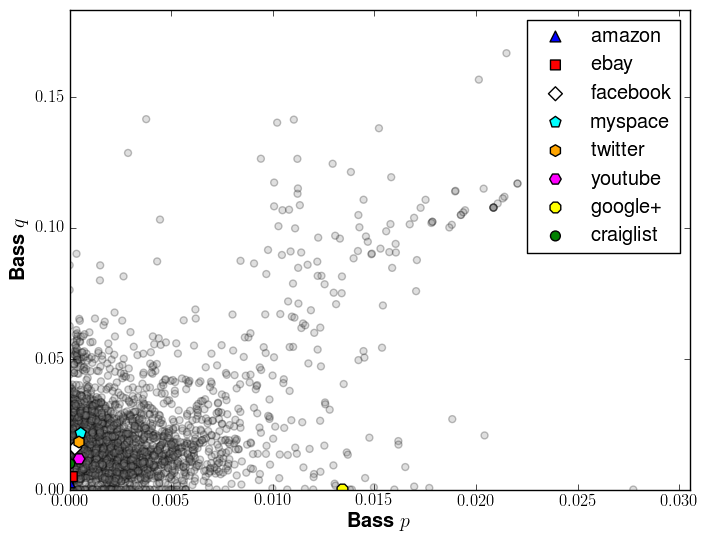}}
\subfigure[shifted Gompertz model]{\includegraphics[width=.32\textwidth]{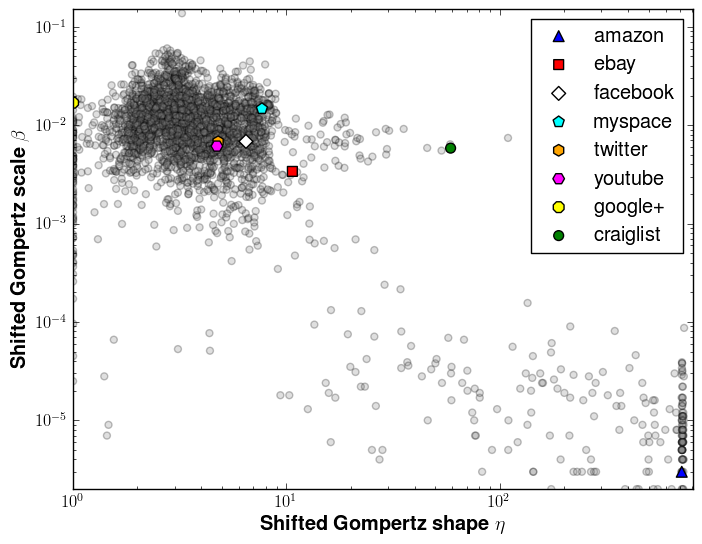}}
\subfigure[Weibull model]{\includegraphics[width=.32\textwidth]{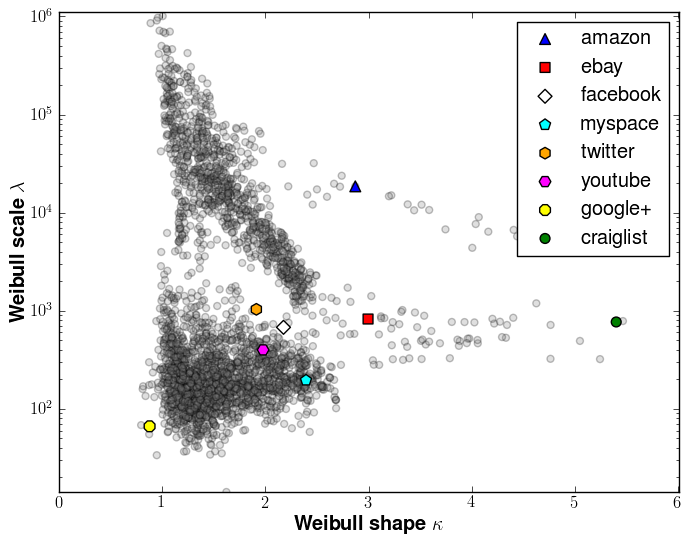}}
\caption{\label{fig:fitembd} Non-linear, two-dimensional embeddings of more than 8.000 search frequency time series into the parameters spaces of the shifted Gomperts-, the Bass- and the Weibull diffusion model. In each case, the 2D embedding coordinates of the eight examples in Fig.~\ref{fig:fitcmp} are highlighted in color.}
\end{figure*}

Figure~\ref{fig:fitcmp} shows how our diffusion models fit general trends for several well known social media platforms and Web-based businesses. Gray curves show  evolving global search volumes available from Google Trends; colored curves represent fitted models where the best fitting one (in terms of GoF) is emphasized. These plots are in line with the results in Tabs.~\ref{tab:gofreg} and \ref{tab:goflang} and illustrate that all three models are able to capture general dynamics even if data for different services show seemingly distinct patterns of growing and declining collective attention.

A considerable advantage of descriptive data mining for attention analysis and, in particular, of using two-parameter diffusion models $f(t \mid \theta_1, \theta_2)$ is that they facilitate \emph{visual analytics}. Once a diffusion model has been fit to a temporal signature of search activities, its parameters $[\theta_1, \theta_2]$ provide as a two-dimensional feature vector that characterizes the time series and may be used in further analysis. Specifically, our approach immediately allows for non-linear, two-dimensional embeddings of the data which can be plotted to visualize whole data sets of time series.

Figure~\ref{fig:fitembd} displays two-dimensional embeddings of all our data according to the different diffusion models. To facilitate interpretation, the coordinates of the eight time series in Fig.~\ref{fig:fitcmp} are highlighted in color.

\begin{figure*}
\centering
\subfigure[US and South Korea]{\includegraphics[width=0.24\textwidth]{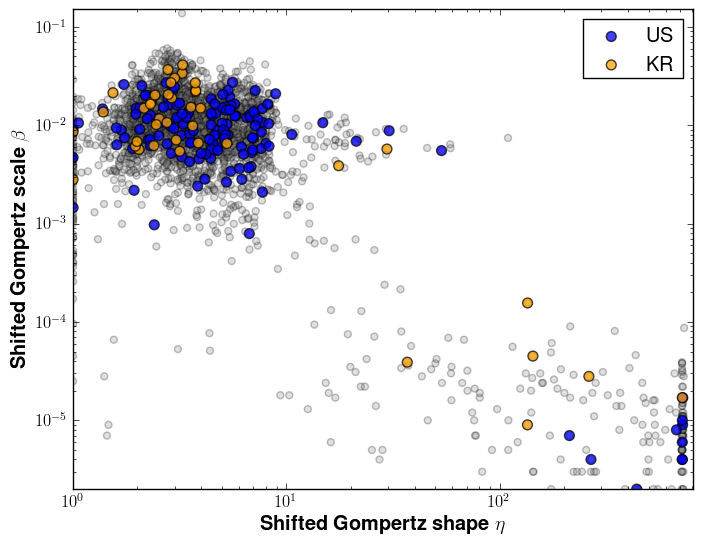}}
\subfigure[Israel and Malaysia]{\includegraphics[width=0.24\textwidth]{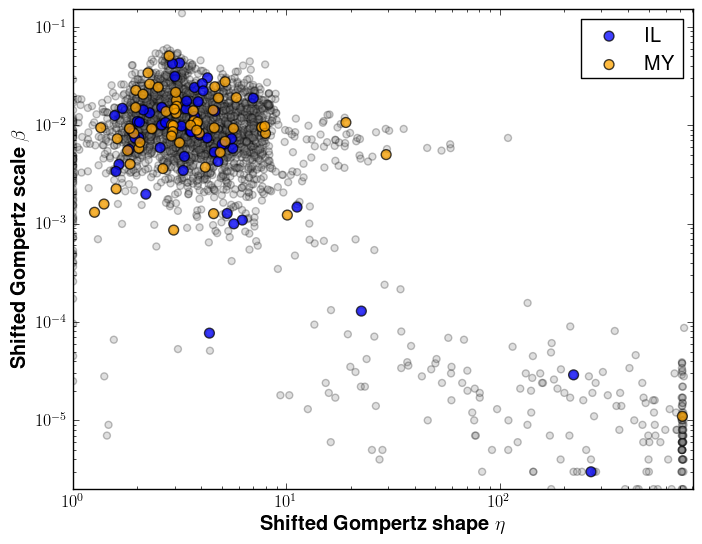}}
\subfigure[Asia and South America]{\includegraphics[width=0.24\textwidth]{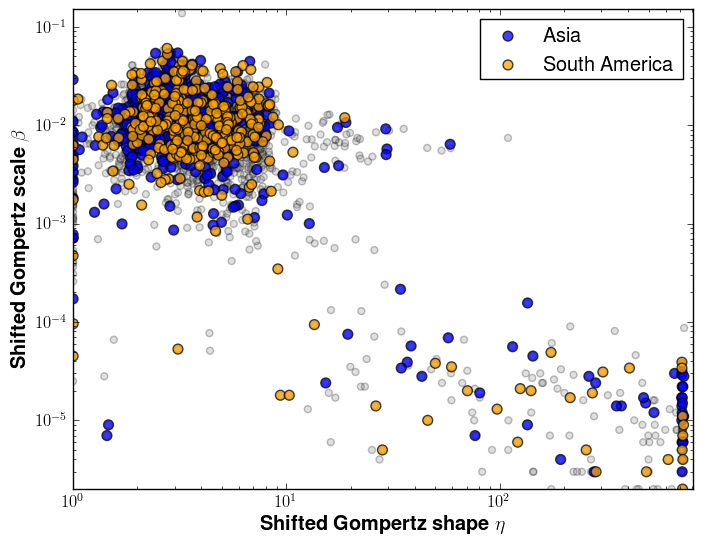}}
\subfigure[English and Russian]{\includegraphics[width=0.24\textwidth]{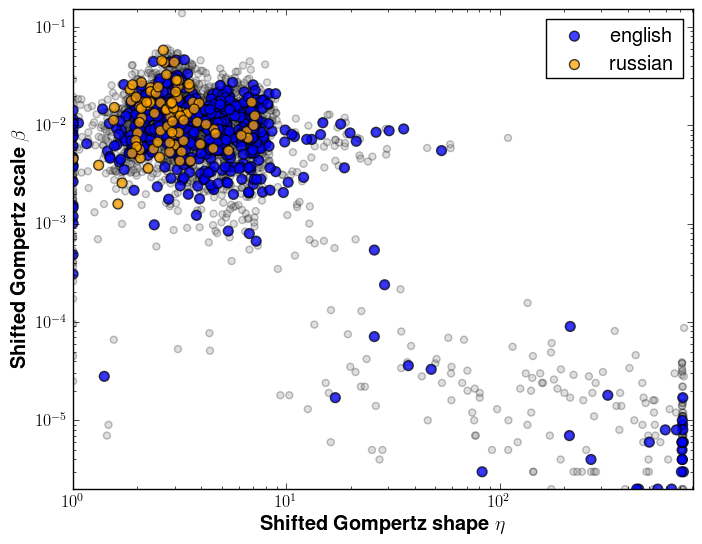}}
\caption{\label{fig:regions} Exemplary comparisons of search frequency time series from different countries, continents, and languages plotted in the two-dimensional parameter space of the shifted Gompertz model.}
\end{figure*}

In each case, the embedding coordinates of \emph{amazon}, a business that continues to attract increasing user interest, marks an extreme location in the embedding space. Similarly extreme locations are occupied by \emph{craiglist} and \emph{ebay}, two Web-platforms that were launched in the 1990s and reached global peak popularity around 2008. The embedding coordinates of \emph{google+}, a service whose search frequency time series indicate a spike of global attention after its launch in 2011, reside at opposite extreme locations. All other time series from Fig.~\ref{fig:fitcmp} are found more or less close together in respective giant clusters of embedded search frequency data.

The existence of these giant clusters which contain almost 90\% of all time series tested is arguably the most important result of our analysis. Irrespective of the diffusion model used to characterize general collective attention dynamics and regardless of which region in the world is considered, it appears that most time series in our collection show similar behavior: \emph{individual social media services seem to be able to attract increasing collective attention for a period of 4 to 6 years before user interest inevitably begins to subside}. This is visible in many of the time series shown throughout this paper, well accounted for by the shape and scale parameters of economic diffusion models, and thus strikingly apparent in Fig.~\ref{fig:fitembd}.

\subsubsection{Case Study: Countries, Continents, Languages}
Figure~\ref{fig:regions} compares examples of attention dynamics for different countries, continents, and linguistic backgrounds.

In Fig.~\ref{fig:regions}(a), we embed data from the US and South Korea in the parameters space of the shifted Gompertz model. Above, both countries were found to be most different regarding median times to adoption of the services considered in this study. Figure~\ref{fig:regions}(a), however, indicates that attention dynamics in both countries are rather similar.

Israel and Malaysia, two countries from different parts of the world, occupy middle ranks in Tab.~\ref{tab:adoption}. Yet, their embeddings in Fig.~\ref{fig:regions}(b) overlap with those of the US and South Korea and do not indicate noteworthy differences as to collective attention dynamics. Similar conclusion apply to the comparison of Asian and South American countries in Fig.~\ref{fig:regions}(c) and the comparison of English and Russian speaking countries in  Fig.~\ref{fig:regions}(d).

\begin{figure}[t]
\centering
\subfigure[\emph{facebook} and \emph{myspace}]{\includegraphics[width=0.23\textwidth]{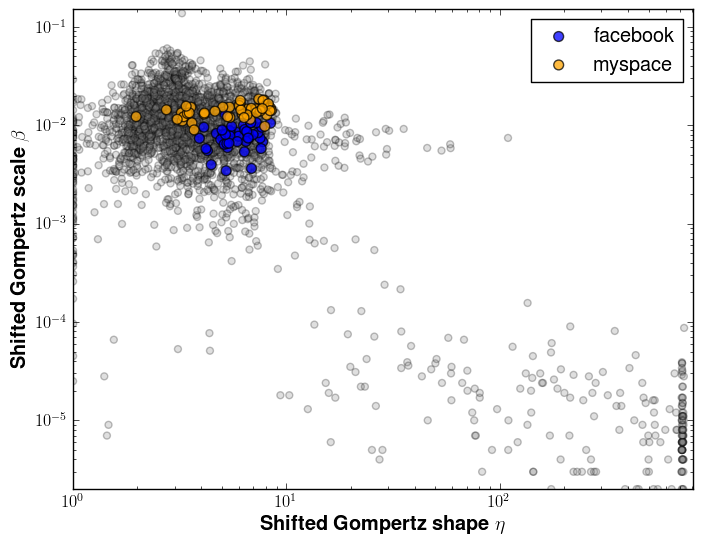}}
\subfigure[\emph{flickr} and \emph{imgur}]{\includegraphics[width=0.23\textwidth]{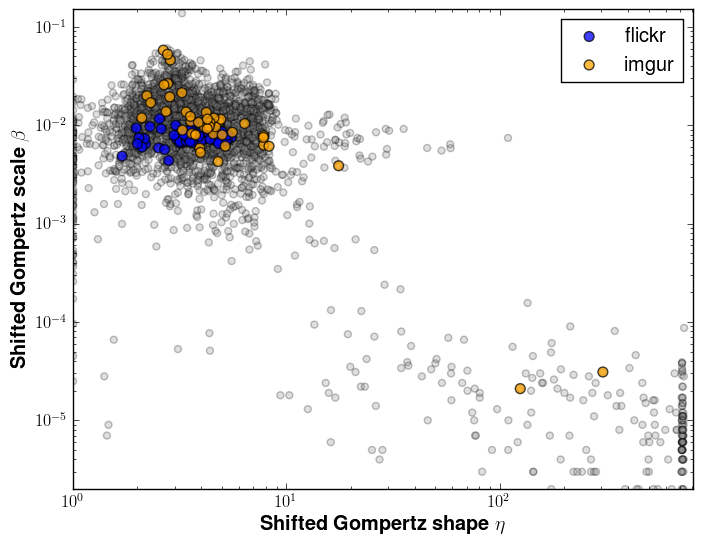}}
\caption{\label{fig:services} Exemplary comparisons of temporal query log data related to different social media services.}
\end{figure}

\subsubsection{Case Study: Social Networks, Photo Sharing}
While diffusion models seem not to allow for a distinction of attention dynamics in different countries or regions, we find that data related to individual services tend to form compact, separable clusters in the parameter spaces of the models we consider.

As an example, Fig.\ref{fig:services} compares two social networks and two photo sharing sites. Country specific time series related to \emph{myspace} and \emph{facebook} form distinct clusters in the embedding space of the shifted Gompertz model. Whereas \emph{myspace} is a social networking site that came and went, \emph{facebook} seems to just have reached global peak popularity. This difference is expressed in the scale parameter of the shifted Gompertz. Likewise, attention dynamics for \emph{flickr} and \emph{imgur} are well explicable in terms of the general cycle of growth and decline; the apparent difference is that, in most countries interest in \emph{flickr} seems to decline while for \emph{imgur} it is still on the rise.

\begin{figure}[t]
\centering
\subfigure[\emph{amazon}]{\includegraphics[width=0.23\textwidth]{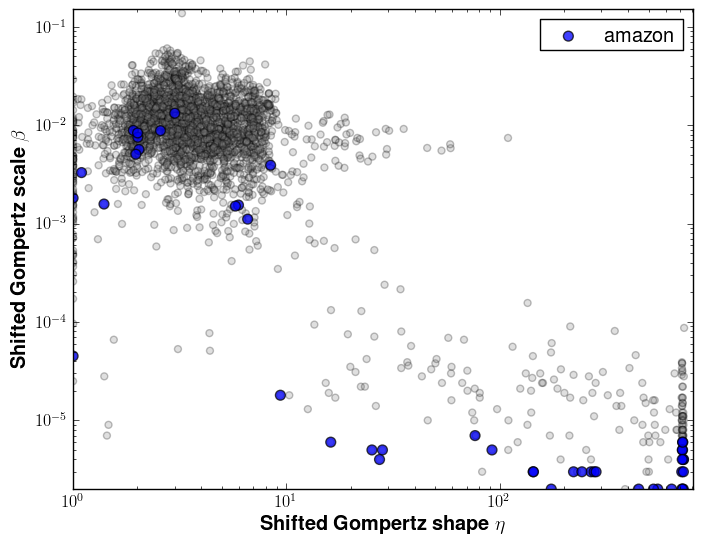}}
\subfigure[UK,DE,VE,FI,JP,ID]{%
\begin{minipage}[b]{0.22\textwidth}
\includegraphics[width=0.49\textwidth]{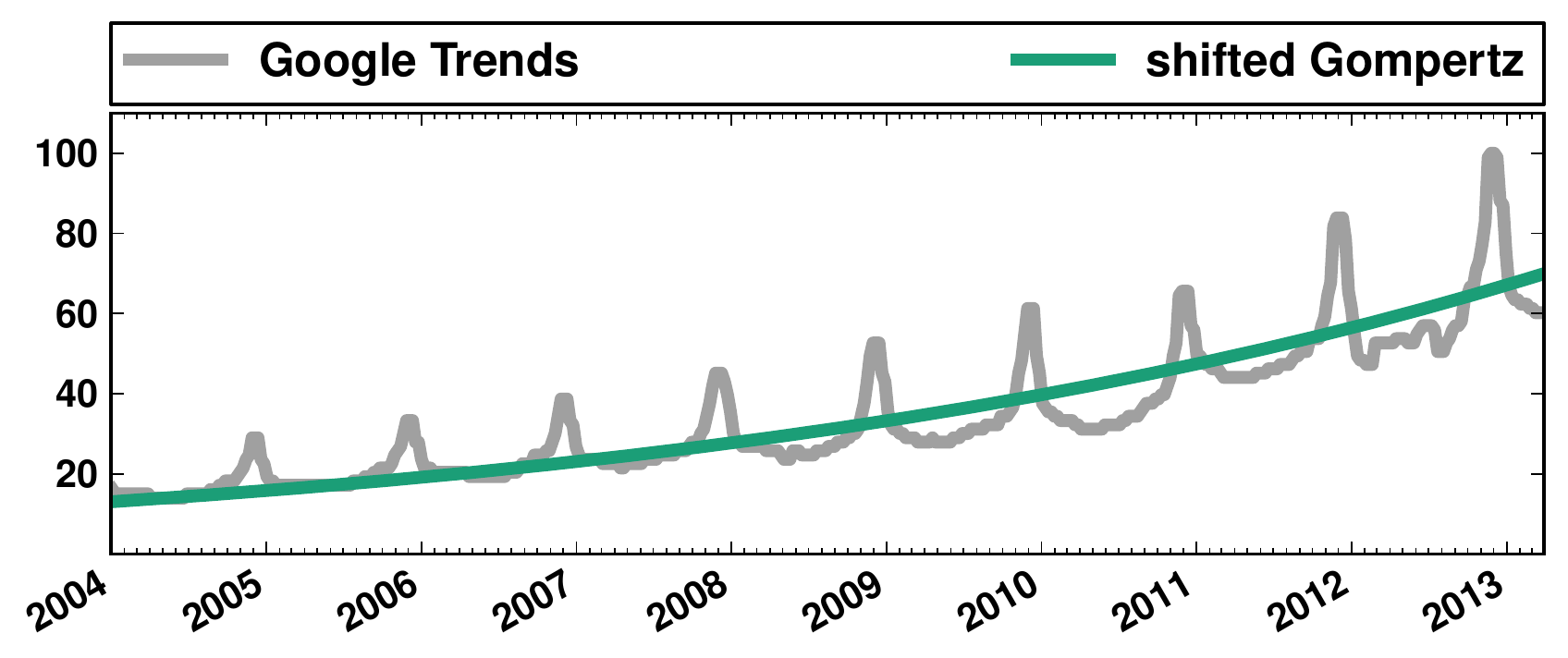}%
\includegraphics[width=0.49\textwidth]{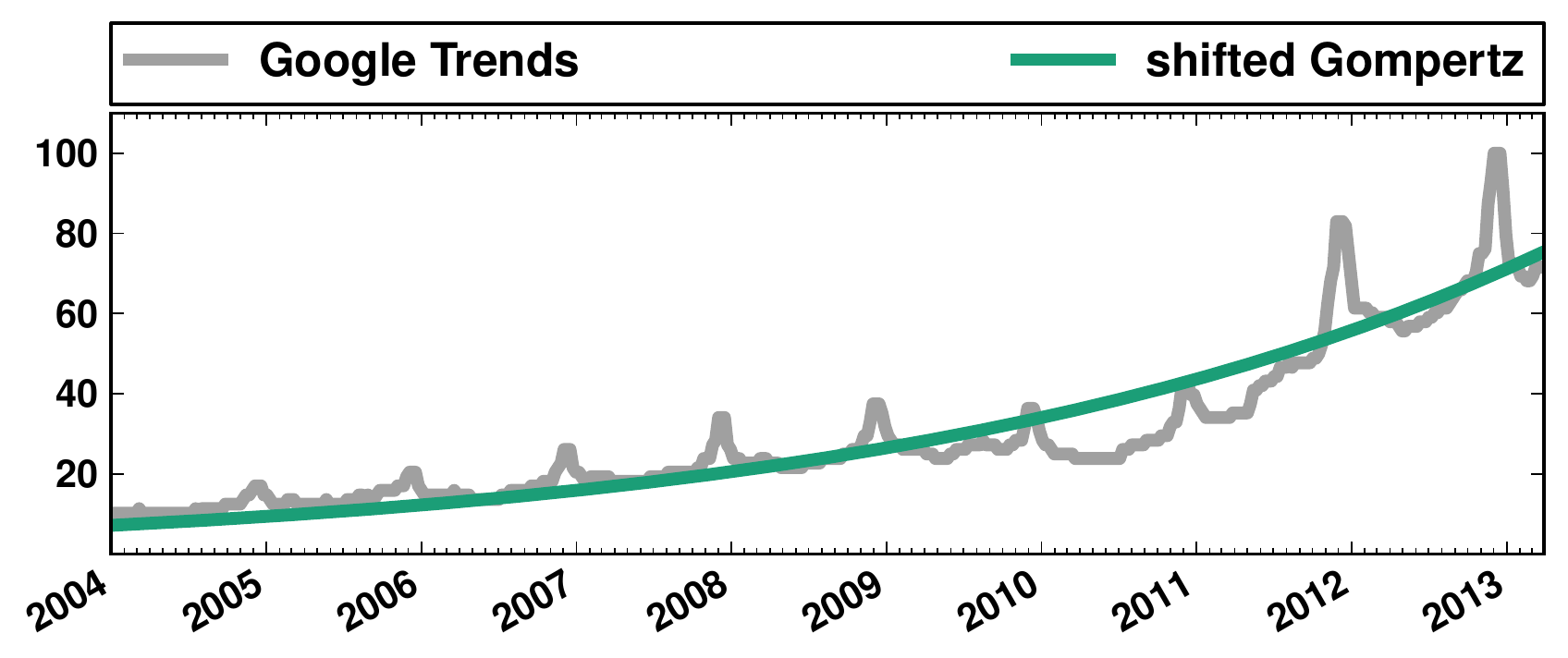}

\includegraphics[width=0.49\textwidth]{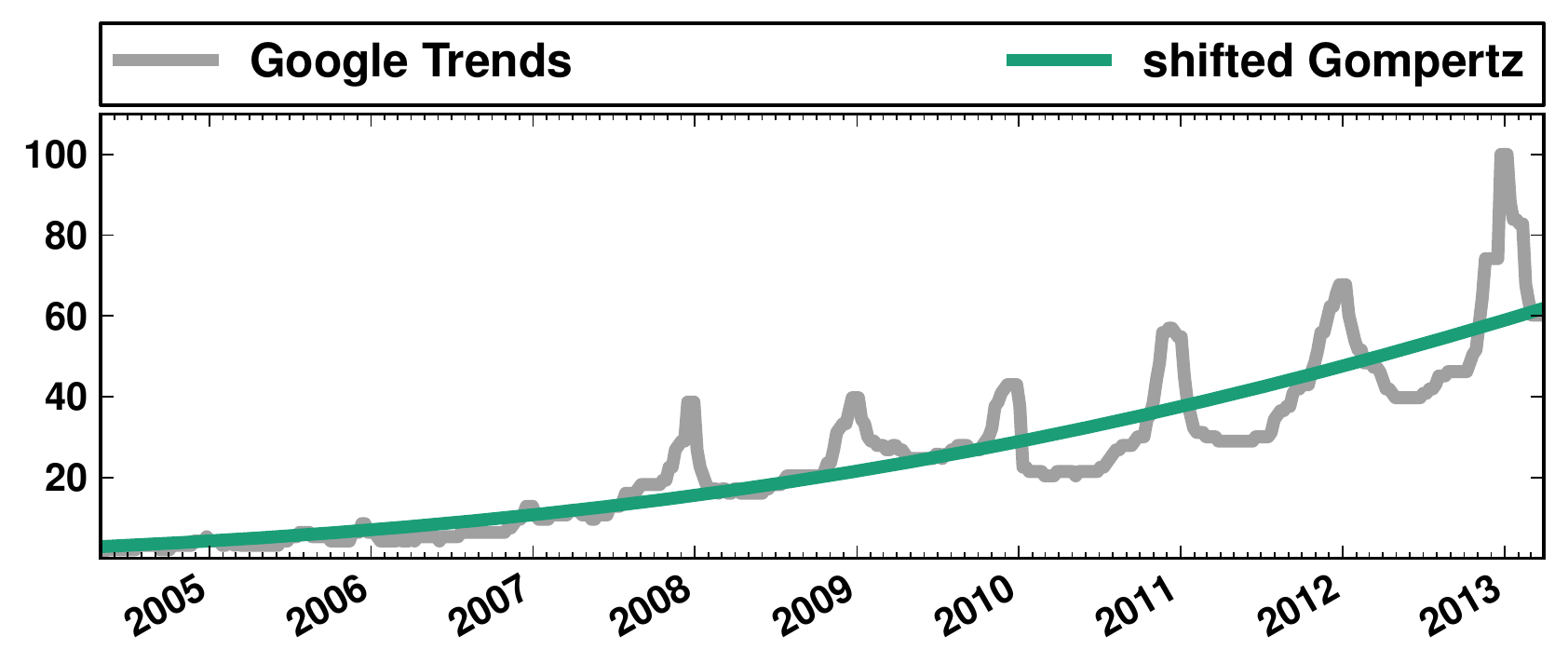}%
\includegraphics[width=0.49\textwidth]{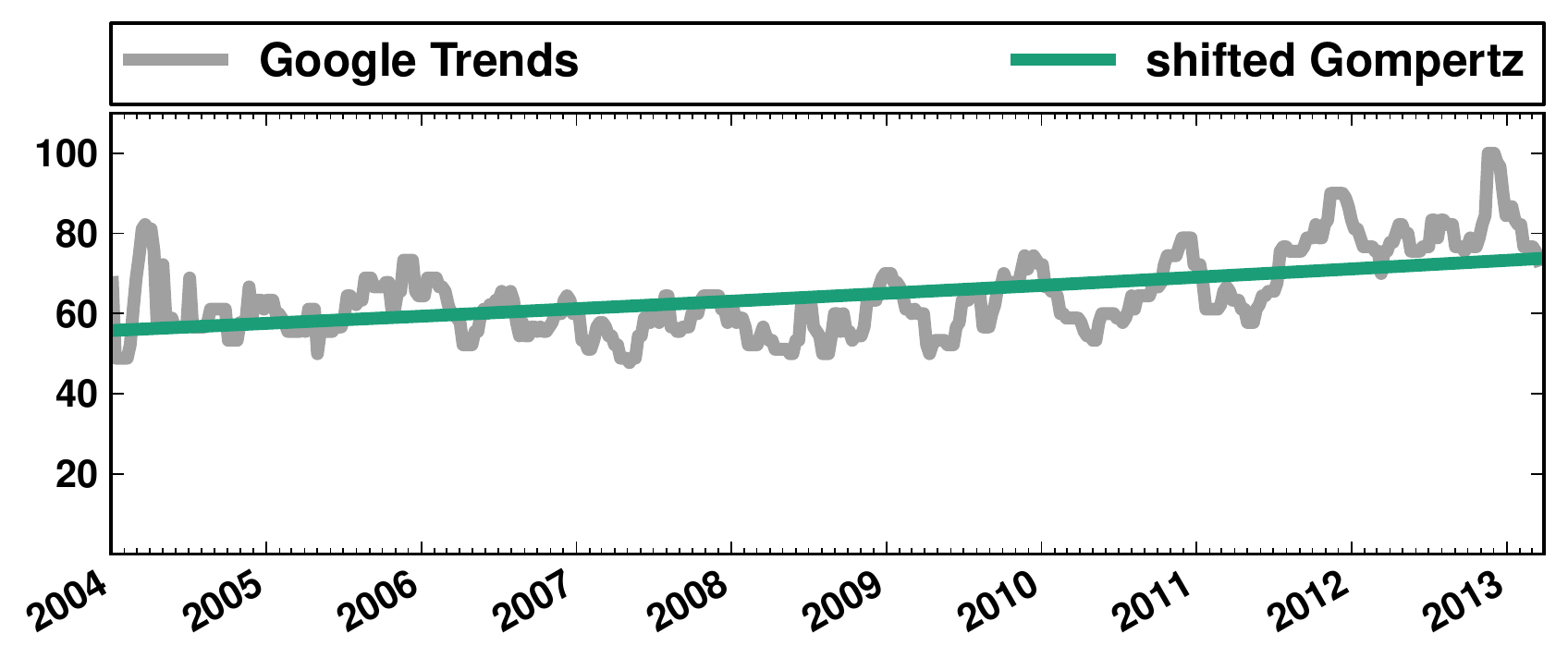}

\includegraphics[width=0.49\textwidth]{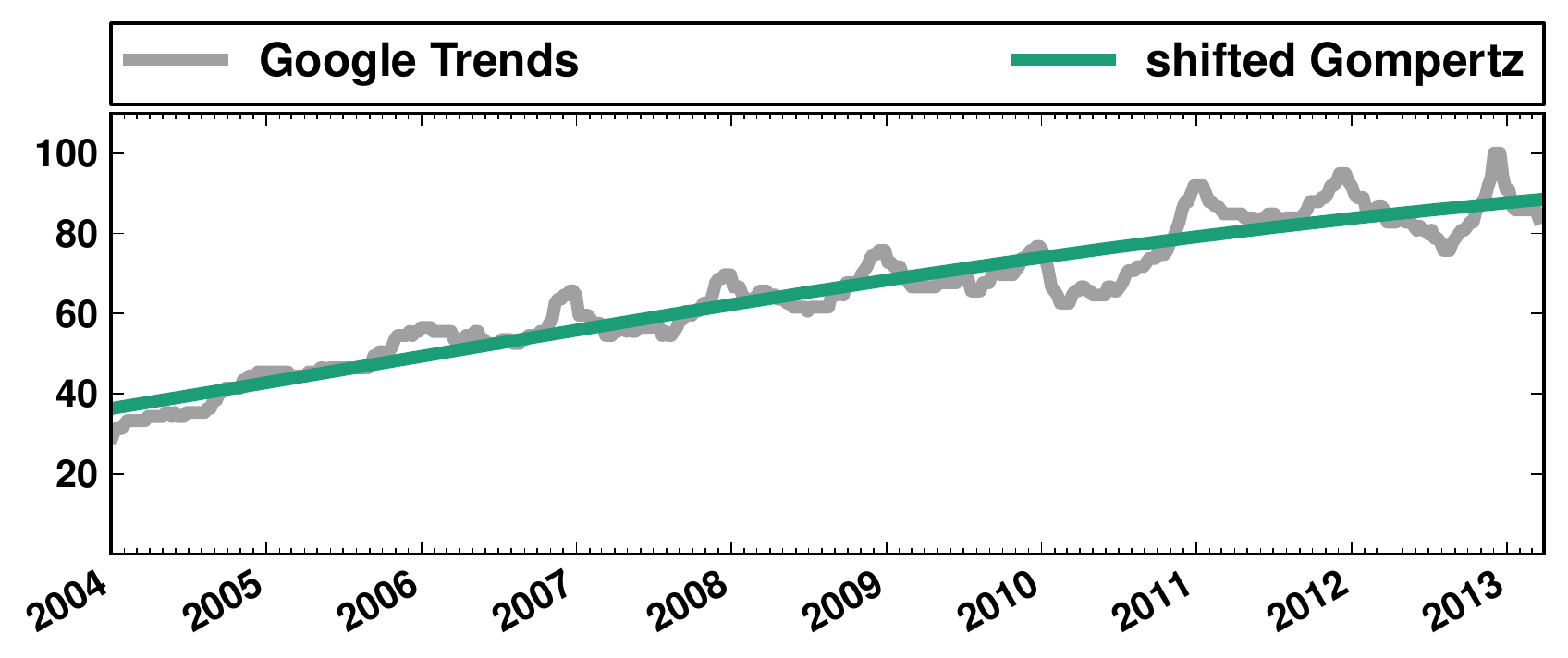}%
\includegraphics[width=0.49\textwidth]{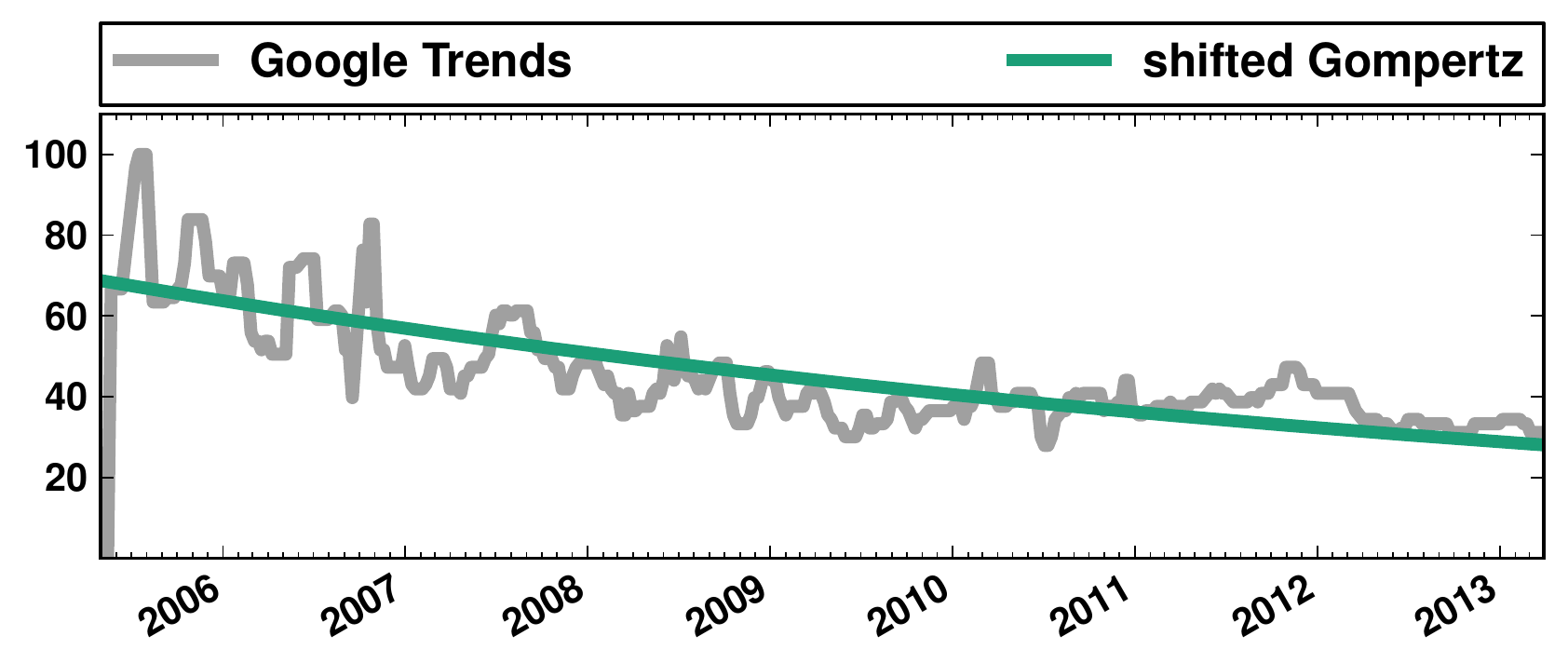}
\vspace{3ex}
\end{minipage}}
\caption{\label{fig:amazon} Query log data related to \emph{amazon}.}
\end{figure}

\subsubsection{Case Study: amazon}
Among all Web-based services considered in this study, \emph{amazon}, an online retailer, is found to cause most diverse patterns of collective attention dynamics.  Figure~\ref{fig:amazon} shows that, while in most countries interest in \emph{amazon} rises steadily over the whole observation period, it remains rather constant in others, and actually declines in some albeit few cases.

\begin{figure}[t]
\centering
\subfigure[\emph{twitter}]{\includegraphics[width=0.23\textwidth]{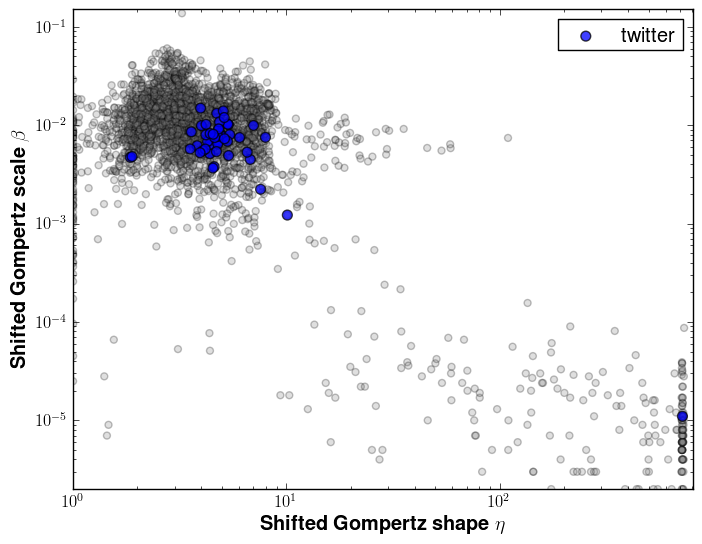}}
\subfigure[NL,PH,RU,FR,MY,TR]{%
\begin{minipage}[b]{0.22\textwidth}
\includegraphics[width=0.49\textwidth]{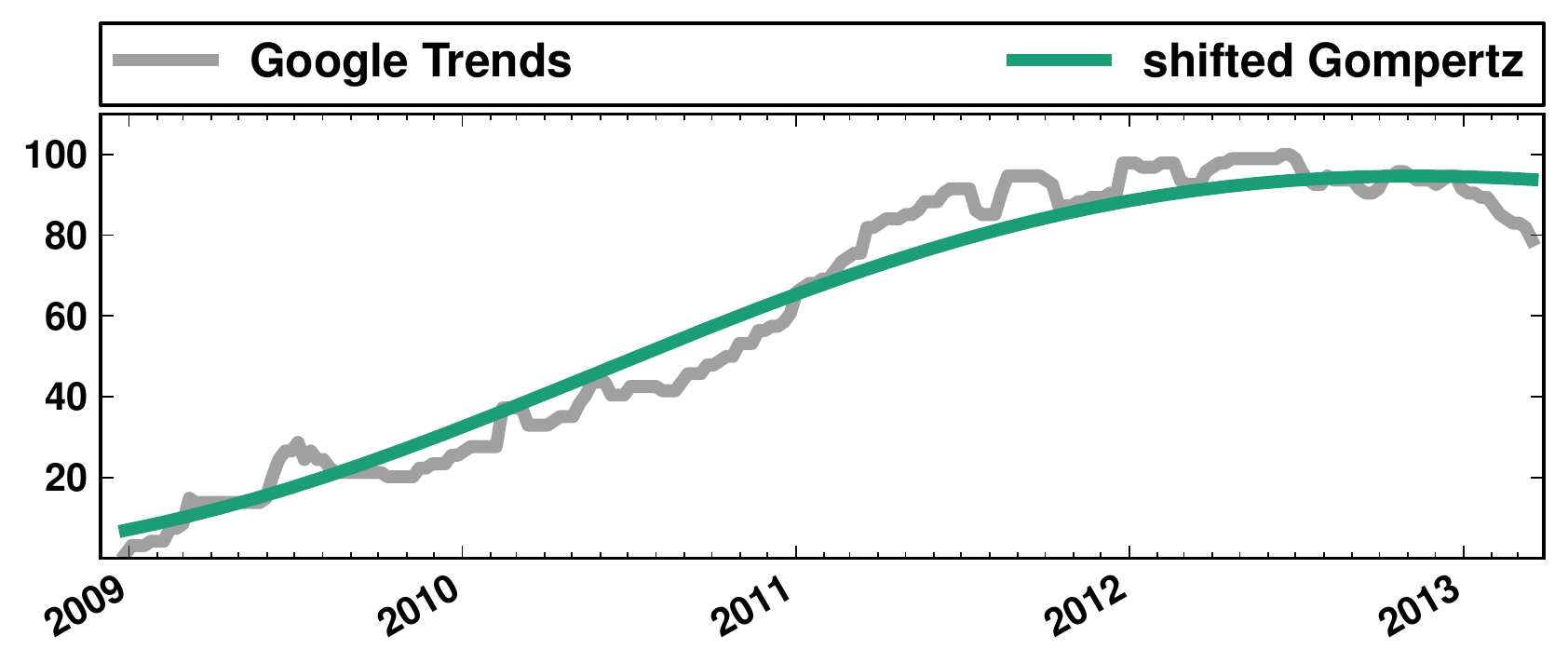}%
\includegraphics[width=0.49\textwidth]{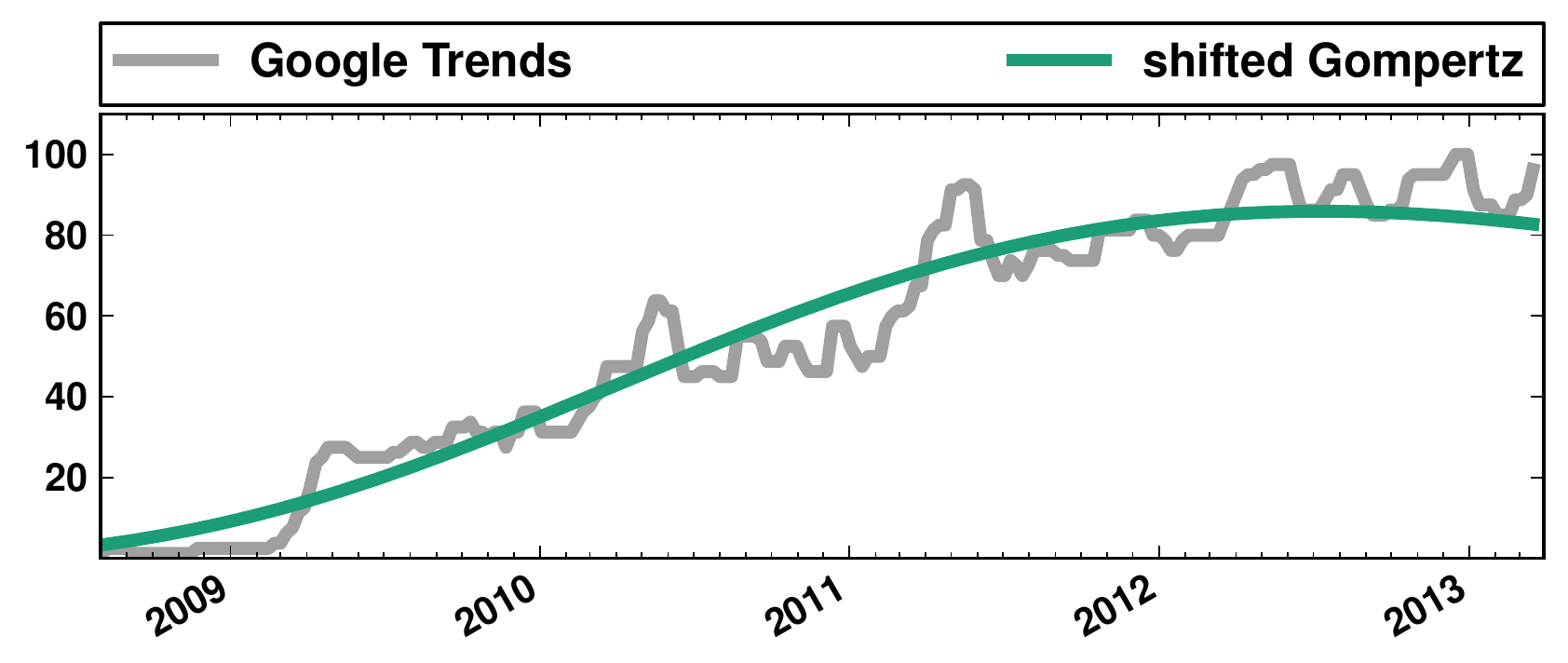}

\includegraphics[width=0.49\textwidth]{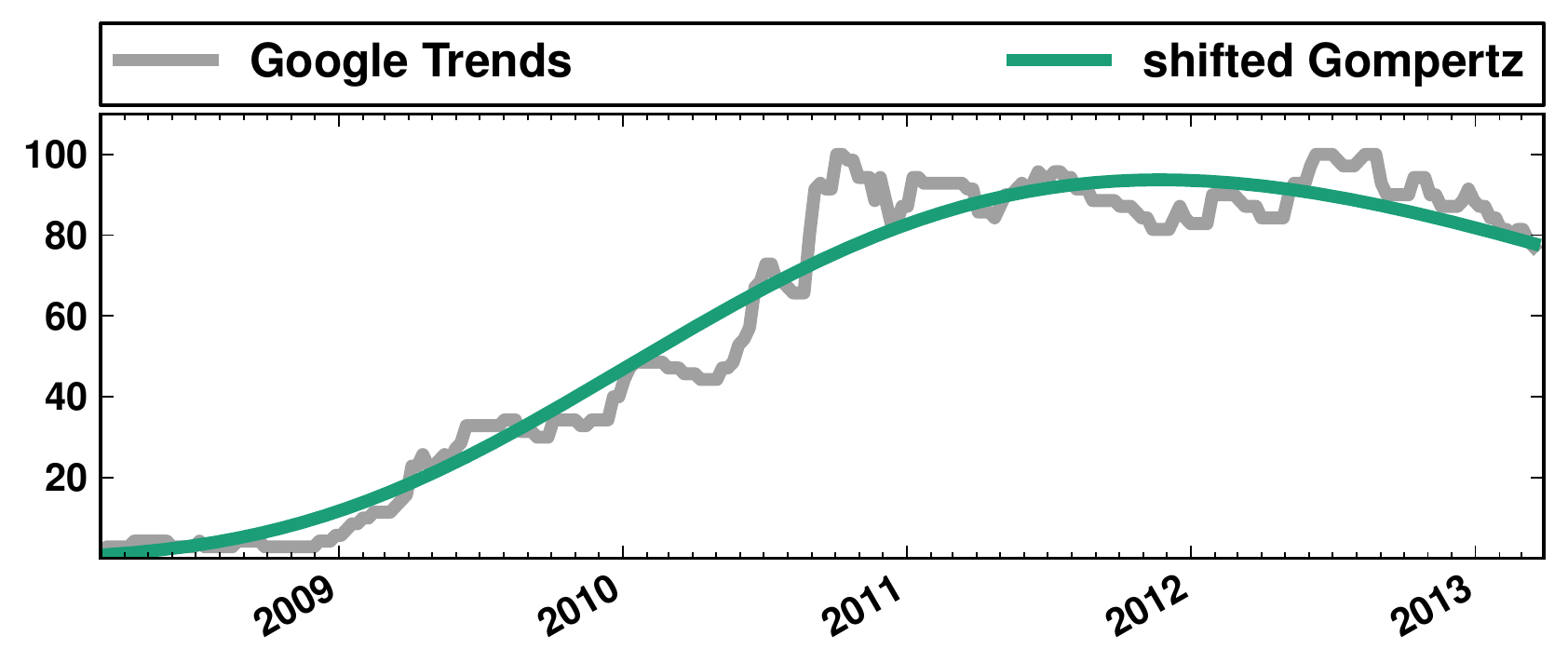}%
\includegraphics[width=0.49\textwidth]{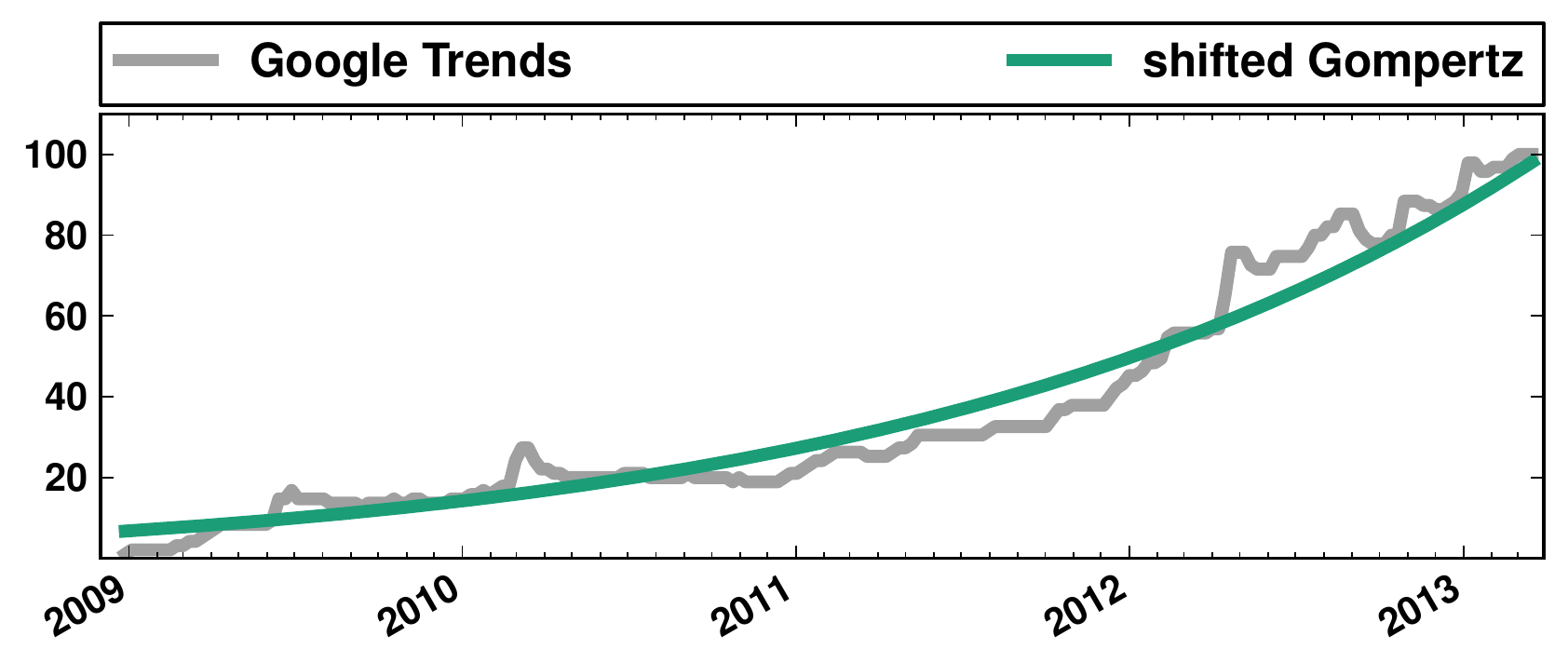}

\includegraphics[width=0.49\textwidth]{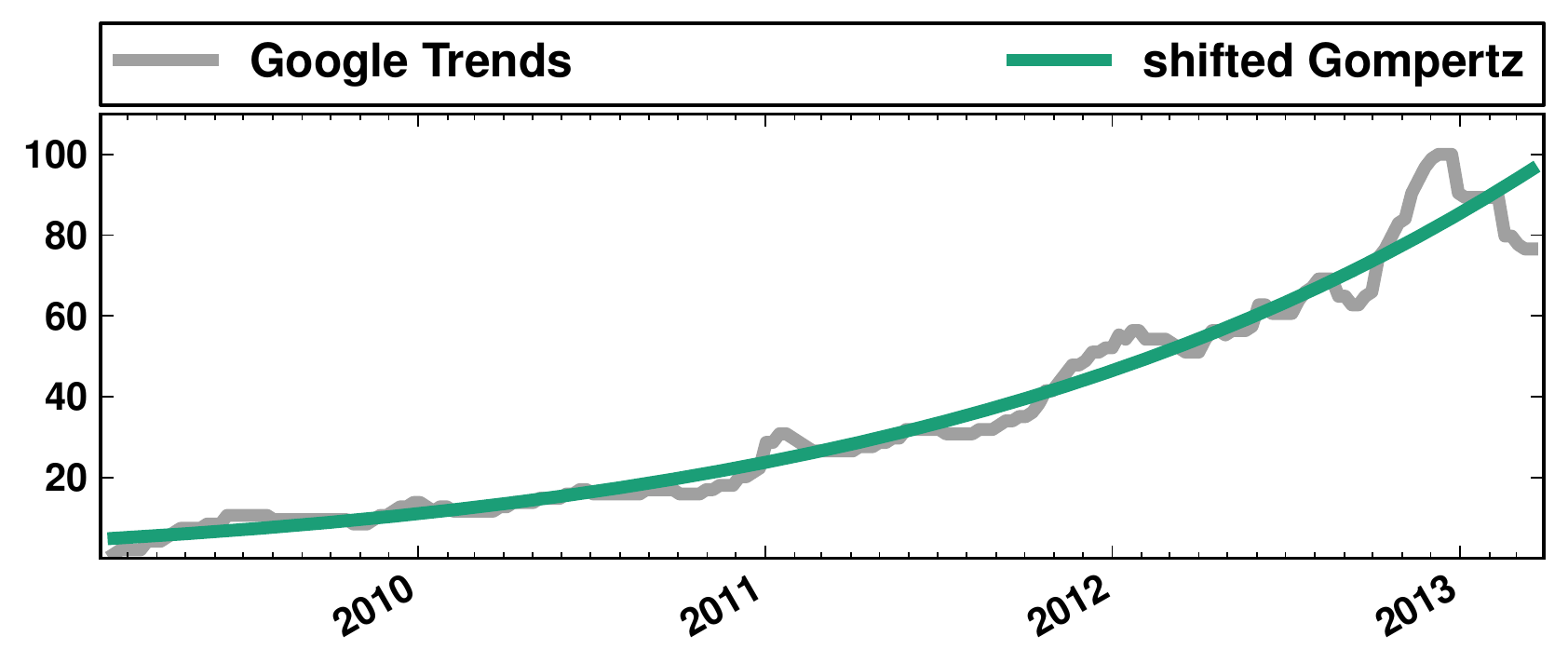}%
\includegraphics[width=0.49\textwidth]{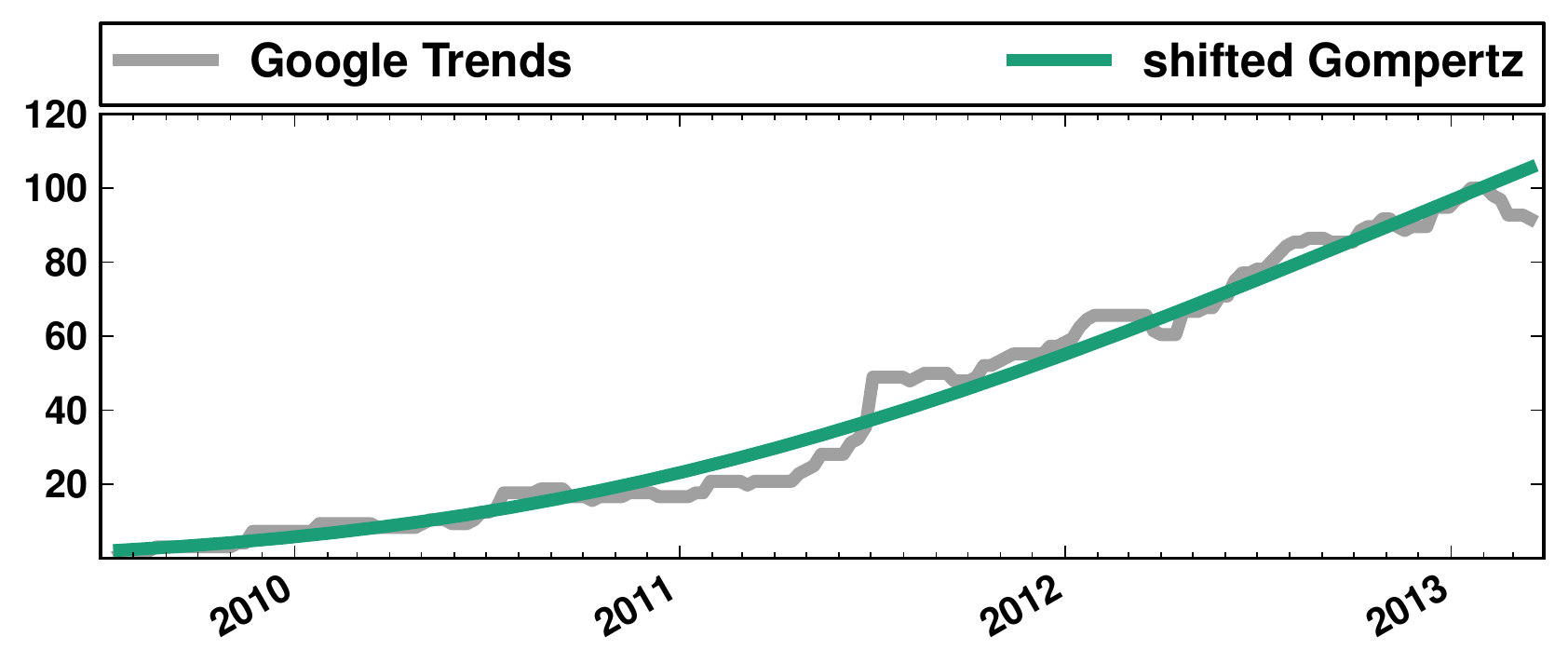}
\vspace{3ex}
\end{minipage}}
\caption{\label{fig:twitter} Query log data related to \emph{twitter}.}
\end{figure}

\subsubsection{Case Study: twitter}
\emph{Twitter}, a popular micro blogging service is another example of a service where attention dynamics vary significantly between countries. While in most countries in our study, interest in \emph{twitter} seems to just have reached its peak (see the distinct cluster within the giant component in Fig.~\ref{fig:twitter}), there are a few countries in which interest in this service continues to rise, notably in France, Malaysia, and Turkey.

\begin{figure*}[t!]
\centering
\subfigure[facebook]{\includegraphics[width=0.32\textwidth]{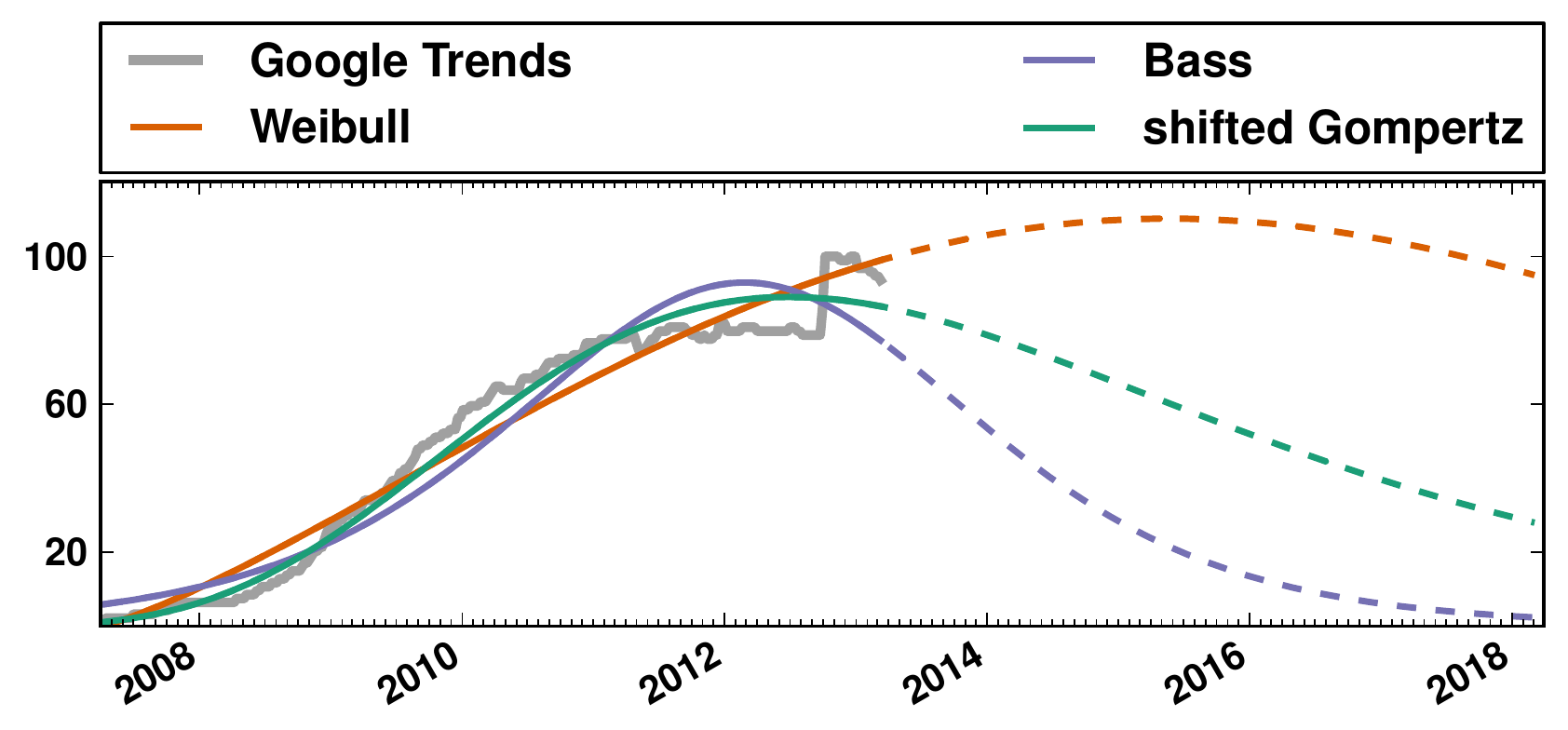}}
\subfigure[youtube]{\includegraphics[width=0.32\textwidth]{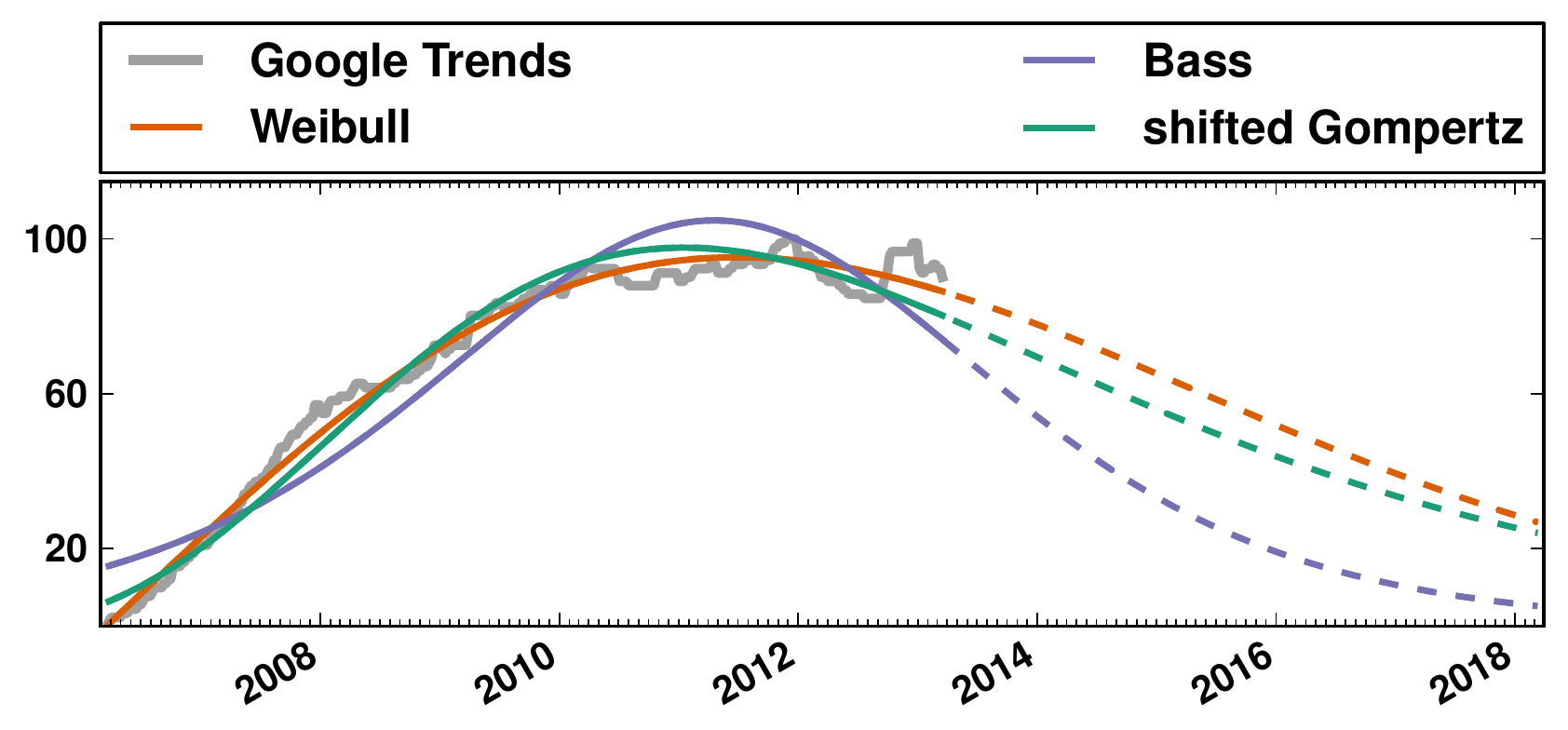}}
\subfigure[twitter]{\includegraphics[width=0.32\textwidth]{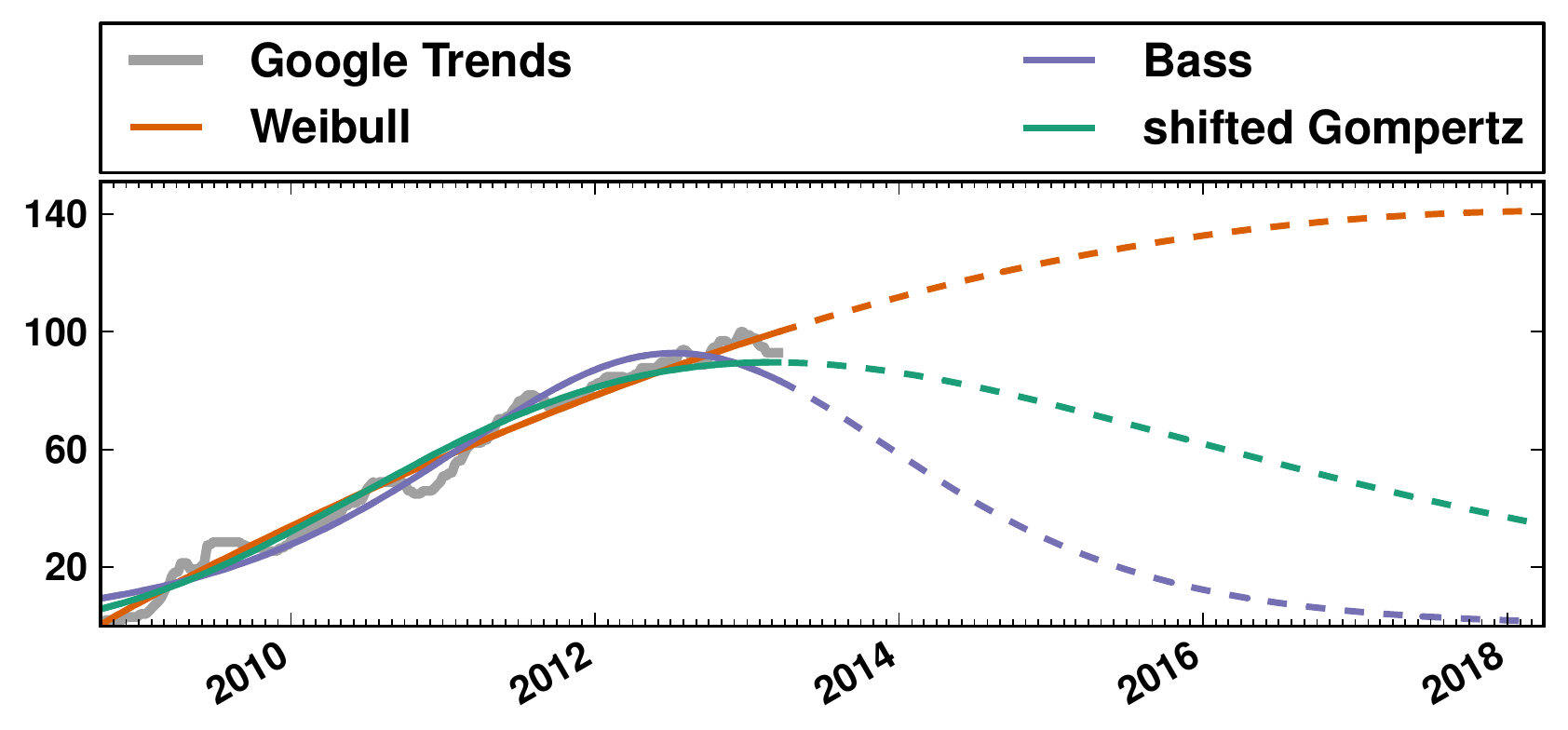}}
\caption{\label{fig:predictions} Predictions of future collective interest in exemplary social media services. Gray curves show data obtained from Google Trends; solid colored curves indicate fits to these data, and dashed colored curves show corresponding 5 year predictions. Note that these predictions do not indicate absolute user interest but predict the evolution of relative search frequencies w.r.t.~the maximum interest so far which is scaled to 100.}
\end{figure*}

\begin{figure*}[t!]
\centering
\subfigure[amazon]{\includegraphics[width=0.32\textwidth]{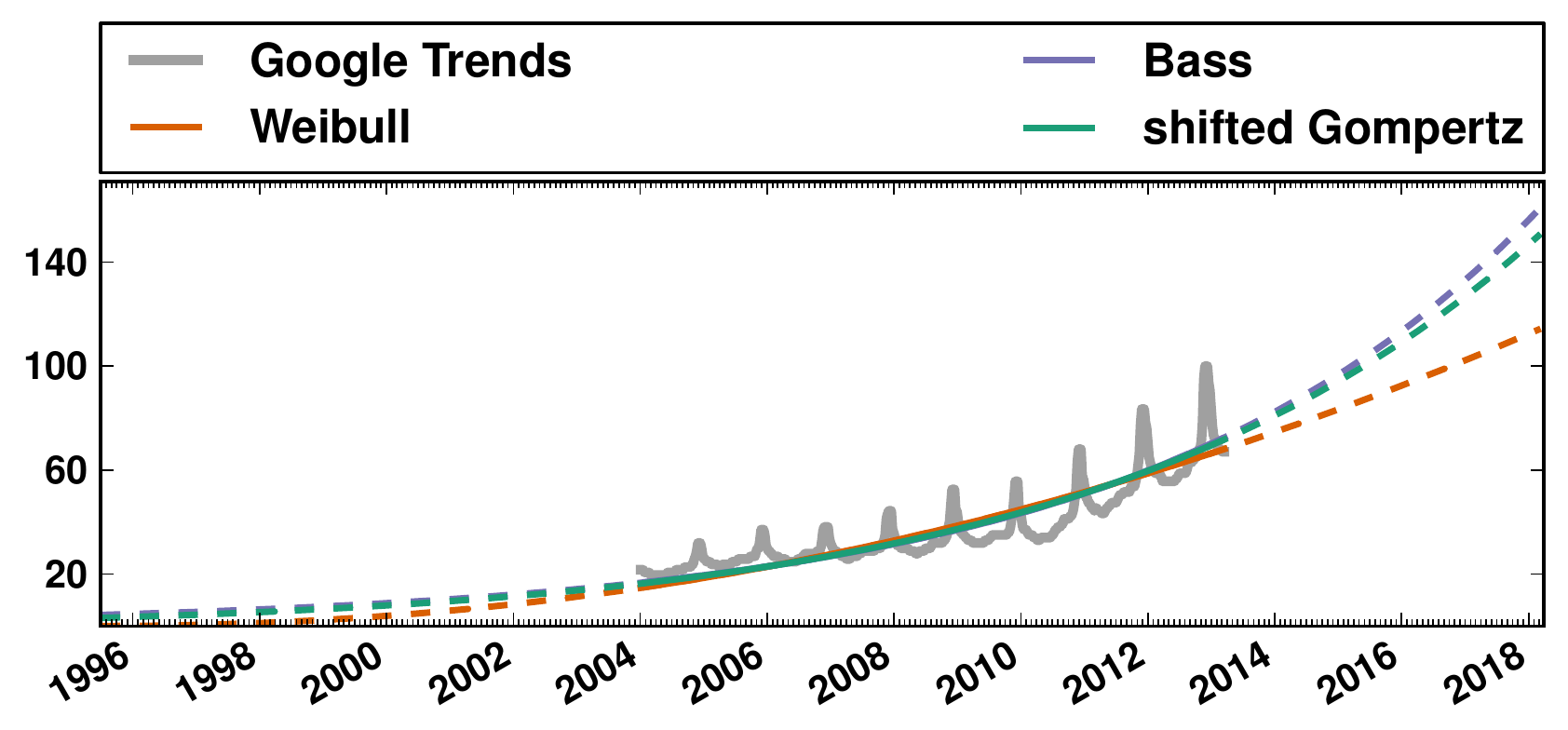}}
\subfigure[paypal]{\includegraphics[width=0.32\textwidth]{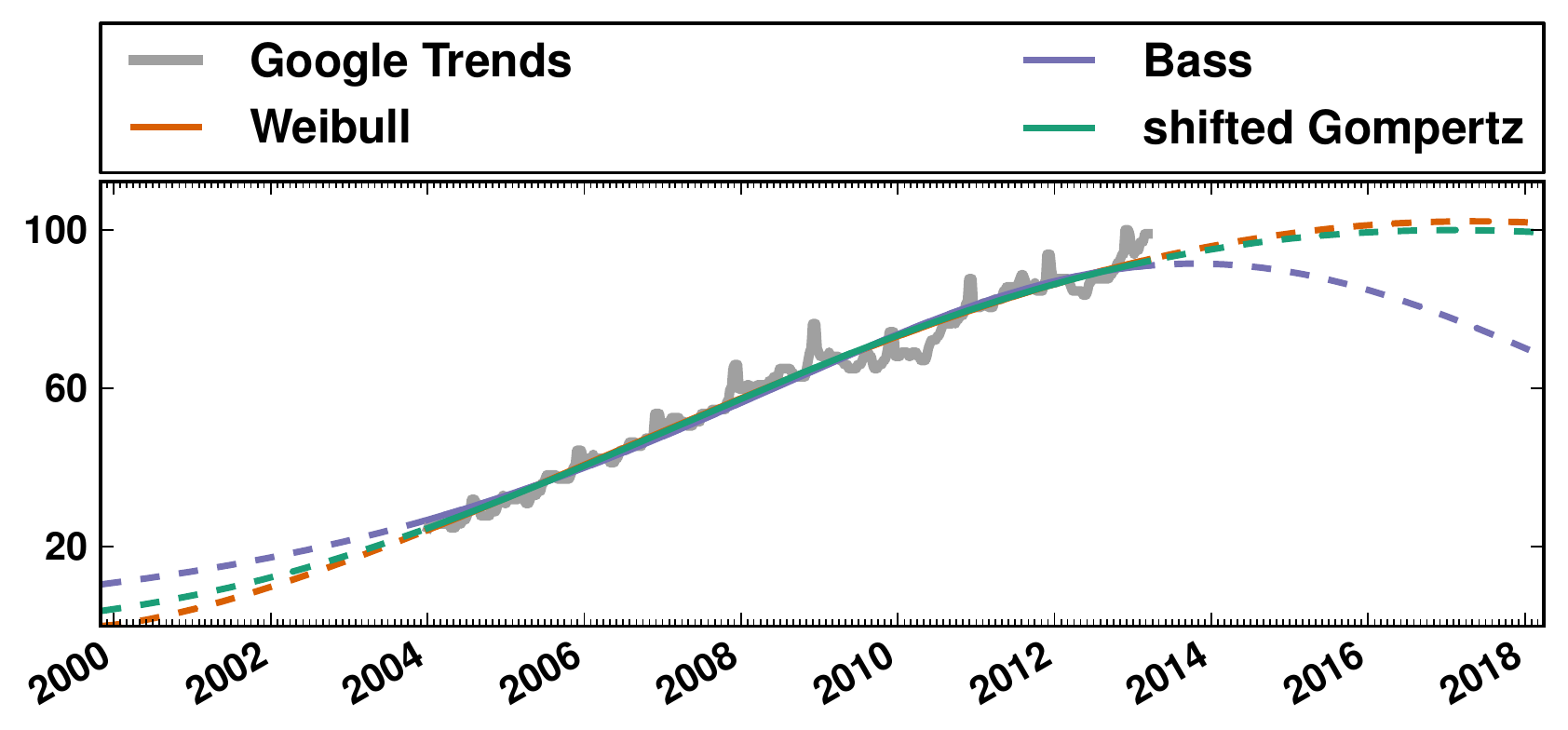}}
\subfigure[ebay]{\includegraphics[width=0.32\textwidth]{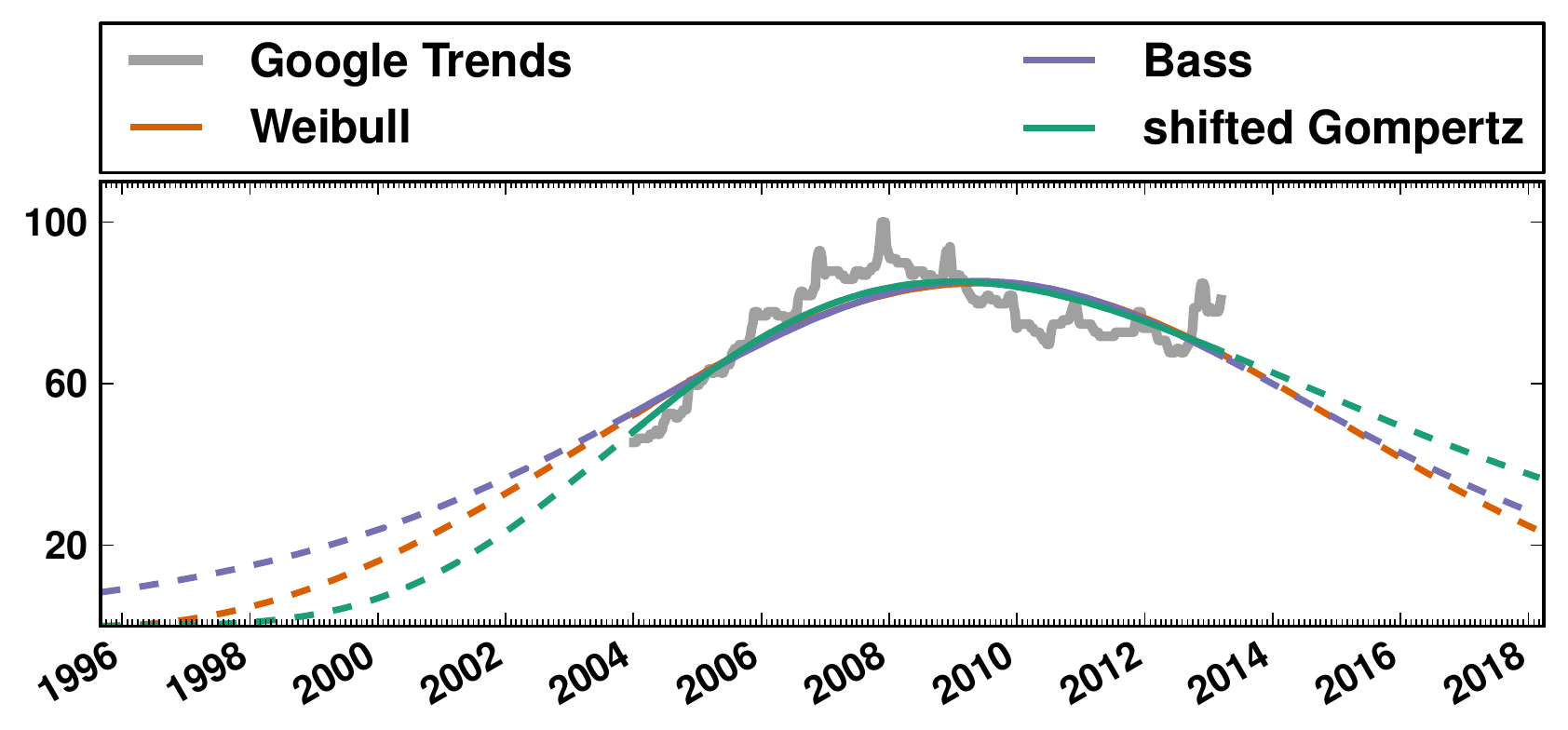}}
\caption{\label{fig:pastdictions} Predictions of past and future collective interest in Web-based businesses launched prior to 2004.}
\end{figure*}

\subsection{Predictions}

Prompted by the overall high statistical significance of fits provided by the three diffusion models, we apply them to predict the future evolution of global collective interest in existing social media. Next, we present qualitative results of predictions over the next five years for exemplary services. In addition, we consider services launched prior to 2004 and demonstrate that the technique discussed in section~\ref{sec:models} also allows for reasonably \emph{predicting the past} and actually is able to reconstruct unobserved past developments.

Figure~\ref{fig:predictions} shows predictions of the development of collective attention to three of todays prominent social media platforms. To create these plots, we scale the range of the best fitting instance of each model in our tests to match the range of values used by Google Trends. While our predictions do not allow for an estimation of the development of absolute numbers of users interested in a service, they indicate how interest may evolving \emph{relative} to the present.

As the available data for the three services is truncated from above, i.e.~each service either has just or has not yet reached peak popularity, traditional maximum likelihood estimates may not be reliable. However, visual inspection suggests that, when using multinomial maximum likelihood, all three diffusion models provide reasonable predictions. While predictions according to the Weibull seem overly optimistic and those due to the Bass model seem rather pessimistic, the shifted Gompertz model marks a middle ground. For instance, in the case of \emph{facebook} it predicts that by 2017 collective interest in this service will reduce to 50\% of its current intensity. We remark that, while at first sight such a development may seem improbable from today's point of view, the vast majority of the 175 social media considered in this paper show characteristic cycles of growth and decline. Given the data available as of this writing, collective attention to \emph{facebook} so far seems to follow the same pattern.

Figure~\ref{fig:pastdictions} shows examples of fits to severely truncated search frequency data. Each Web-based business in this figure was launched prior to 2004 so that data from Google Trends is incomplete regarding the past. Nevertheless, based on multinomial maximum likelihood, the characterizations of general trends according to each diffusion model are again reasonable; in particular, onset times predicted by the shifted Gompertz match the dates these businesses were launched.

\section{Related Work}
\label{sec:relatedwork}

Understanding the collective behavior of crowds of Web users is a research topic of growing popularity and model-based approaches have been used in this context before. We divide our discussion of related work into two major parts: first, we review previous contributions to attention dynamics on the Web in general and then we discuss two recent, highly related publications on the evolution of popularity of social media services which themselves have stirred considerable attention in early 2014.

\subsection{Attention Dynamics on the Web}

Statistical distributions similar to the ones considered in this paper have been previously applied to characterize the dynamics of the behavior ow crowds of Web users.
In an early contribution, Huberman et al.~\cite{Huberman1998-SRI} analyzed browsing behaviors and found that the number of links a user is likely to follow on a Web site is distributed according to an inverse Gaussian. In \cite{Wu2007-NAC}, Wu and Huberman studied life-cycles of news items on social bookmarking site and found that the amount of attention novel content receives is distributed log-normally. The log-Normal distribution was also found to model sizes of cascades of messages passed through a peer-to-peer recommendation network \cite{Leskovec2007-TDO} or the number of messages exchanged in instant messaging services \cite{Lescovec2008-PSV}.

The Weibull distribution in \eqref{eq:Weibull} was recently reported to account well for statistics of dwell times on Web sites \cite{Liu2010-UWB}, times people spend playing online games \cite{Bauckhage2012-HPL}, or the dynamics of Internet memes \cite{Bauckhage2013-MMO}. The Bass diffusion model in \eqref{eq:BassHazard} has recently been considered in order to reason about structures of online social networks \cite{Luu2012-MDI} or \emph {twitter} information cascades \cite{Hermann2013-AAB}. The shifted Gompertz distribution, on the other hand, was apparently not yet considered in the context of social media or Web usage dynamics.

While attention dynamics on shorter time scales have been modeled using random fields \cite{Lin2010-PET}, structured models \cite{Goel2012-TSO}, or differential equations \cite{Leskovec2009-MTA}, long term temporal dynamics of collective attention have previously been modeled using mixtures of power-law and Poisson distributions \cite{Crane2008-RDC} or systems of differential equations \cite{Acerbi2012-TLO,Leskovec2007-TDO} which were inspired by techniques from the area of epidemic modeling \cite{Britton2010-SEM,Dietz1967-EAR}. In this context, we note that the diffusion models considered in this paper also allow for interpretations in terms of the dynamics of elementary differential equations. For instance, the Weibull model in \eqref{eq:Weibull} can be expressed as $f(t) = \frac{d}{dt} F(t) = \alpha \kappa t^{\kappa-1} - \alpha \kappa t^{\kappa-1} F(t)$ which hints at a similarity in spirit between economic diffusion and established epidemic models that seems to merit further research.

With respect to time series retrieved from Google Trends, epidemic models based on differential equations involving exogenous end endogenous influences have been discussed in
\cite{Crane2008-RDC}. There, they were used as means of classifying, i.e.~distinguishing, different types of attention dynamics. Trend analysis based on data from Google Trends was also performed in \cite{Christiansen2012-MTT} yet there the focus was on developing clustering algorithms to characterize different phases in search frequency data. The approaches in \cite{Christiansen2012-MTT,Crane2008-RDC} are thus related to what is reported here, however, in contrast to these contributions, we do not explicitly devise new models but consider simpler representations that implicitly account for different kinds of dynamics. Due to the simplicity of the diffusion models considered here and because of their apparent empirical validity and theoretical plausibility, the results reported in this paper therefore provide a new baseline for research on the mechanisms and long-term dynamics of collective attention on the Web.

\subsection{The Princeton / Facebook Controversy and a Contribution from CMU}

In a delightful synchronicity, Cannarella and Spechler \cite{Cannarella2014-EMO}, Ribeiro \cite{Ribeiro2014-MAP}, and we ourselves \cite{Bauckhage2014-CAT} all published analyzes on how attention to social media evolves over time in early 2014. While \cite{Cannarella2014-EMO} was uploaded to arXiv, \cite{Ribeiro2014-MAP} and \cite{Cannarella2014-EMO} were both presented at the International World Wide Web Conference in Seoul.

The work by Cannarella and Spechler from Princeton is noteworthy for triggering a brief but fierce media frenzy. Just as in the work presented here, the results in \cite{Cannarella2014-EMO} were obtained from analyzing Google Trends time series. Differing from our approach, Cannarella and Spechler considered epidemic models to analyze search frequency time series that indicate interest in services such as \emph{myspace} or \emph{facebook}. While this methodology had earlier been applied to analyze the temporal evolution of interest in Internet memes \cite{Bauckhage2011-III}, Cannarella and Spechler caused a controversy, because they used their models to predict that \emph{facebook} would lose 80\% of their users by 2017. Media interst was particularly stirred by the fact that \emph{facebook} data scientist Mike Develin was quick to humorously ``debunk'' the Princeton ``findings''\footnote{see: www.facebook.com/notes/mike-develin/debunking-princeton/10151947421191849}.

Interestingly, our ``qualitative'' results in Fig.~\ref{fig:predictions} seem to corroborate Cannarella's and Spechler's predictions and we note that they were obtained from the same data but different models. In any case, we certainly agree with Develin's objection that predictions based on search frequency data have to be taken with a grain of salt. Yet, we disagree with his argument that social media related search interests of millions of Web users are not indicative of user engagement (see again our discussion in Section~\ref{sec:google}) and note the curious absence of any direct engagement data in his reply.

However, data that directly reflects engagement played an important role in Ribeiro's analysis performed at CMU \cite{Ribeiro2014-MAP}. He considered statistics available from \emph{alexa}, a subsidiary of \emph{amazon} which provides Web traffic data that are gathered using the \emph{alexa} toolbar, a plugin that volunteers install in their browsers so that \emph{alexa} can track which Web pages they access.

Regarding Ribeiro's approach, we note that he extended established epidemic models by new parameters and found these new models to be in good agreement with his data. His findings, too, caused considerable media attention since he predicted collective user interest in \emph{facebook} to remain constant for years to come. Yet, this result as well should be taken with a grain of salt. While it was derived from direct engagement data, we point out that \emph{alexa} data are likely biased towards technology savvy users who installed the toolbar and will hardly reflect the surfing behavior of average Web users.

Given this discussion, the approach and results presented here mark a middle ground. On the one hand, we consider simple diffusion models rather than (intricate) models for the epidemic spread of novelties. On the other hand, the statistical basis for our analysis far exceeds those in \cite{Cannarella2014-EMO,Ribeiro2014-MAP}. Neither  Cannarella and Spechler nor Ribeiro consider country specific data and neither of them considers as large a number of different services than we do in this paper. Moreover, we see the main contribution of this paper not in the predictions in Fig.~\ref{fig:predictions} but rather in the empirical observation that collective attention to social media shows highly regular patterns of growth and decline regardless of region of origin or cultural background of crowds of Web users.

\section{Conclusion}
\label{sec:conclusion}

In this paper, we performed search frequency analysis in order to gain insights into the dynamics of collective attention to social media and Web-based businesses. Search frequency analysis is an emerging topic and a quickly growing literature shows that data available from Google Trends can lead to novel insights into collective concerns, interests, or habits \cite{Artola2012-TTF,Bauckhage2011-III,Bauckhage2013-MMO,Bordino2013-WSQ,Castle2009-NIN,Choi2012-PTP,Da2011-ISO,Gerow2011-MTW,Granka2009-ITP,Joseph2011-FAS,McLaren2011-UIS,Mellon2011-SIA,Teevan2011-UAP}.

Interested in collective attention to social media, we collected Google Trends data from 45 different countries that show how user interests in 175 social media services evolved over time. Focusing on general trends, we considered descriptive data mining techniques and applied economic diffusion models to search our data set of more than 8,000 times series for common patterns or distinctive differences.

Diffusion models are well established in economics and we considered their use due to their conceptual simplicity. This is in contrast to more elaborate approaches such as, say, Gaussian mixtures or kernel techniques, which yield results in terms of parameters for which there usually is no physically plausible counterpart. Diffusion models, on the other hand, are designed to characterize time series in terms of everyday concepts such as propensities for attention to grow and to decline and we note that Occam's razor suggests to prefer simple explanations whenever available.

Using an efficient algorithm for robust maximum likelihood parameter estimation even under incomplete data, we fitted the Bass-, the shifted Gompertz-, and the Weibull diffusion model and evaluated their performance. Our most important results can be summarized as follows:
\begin{itemize}
\item economic diffusion models provide accurate and statistically significant explanations of general trends in aggregated search frequency data which summarize how collective attention to social media evolves over time.
\end{itemize}
This capability of diffusion models to characterize the data considered in this study thus suggests that:
\begin{itemize}
\item collective attention to social media evolves according to simple and highly regular dynamics of growth and decline.
\end{itemize}
In a comparative analysis w.r.t.~individual countries, different continents, or linguistic backgrounds, we found these patterns to be persistent and conclude that
\begin{itemize}
\item collective attention to social media evolves globally similarly and independent of regions of origin or cultural backgrounds of crowds of Web users.
\end{itemize}
Regarding individual services, however, rates of adoption may vary between countries. Nevertheless, for almost 90\% of the time series in our data set, we found strikingly similar attention dynamics and it seems that
\begin{itemize}
\item most social media services are able to attract growing collective attention for a period of 4 to 6 years before user interest inevitably begins to subside.
\end{itemize}

Finally, because of the way growth dynamics are encoded in the diffusion models studied here, it appears that public attention to social media hinges on  perceived novelty. In other words, the more a crowd of users gets used to a service or the less novel it appears, the faster it looses its appeal. These are the characteristics of hype cycles. The temporal behavior exposed in our analysis is therefore well in line with everyday experience and aptly summarized by the statement that what goes up, must come down.

Our results are of interest to professionals in marketing and public relations. According to findings in \cite{Mellon2011-SIA,Teevan2011-UAP} pertaining to the saliency of query logs for behavioral studies, data which aggregate the Web search behavior of millions of people worldwide provide reasonable proxies for public interests and preferences. The strongly regular patterns we identified in time series that served as proxies for the popularity of social media therefore indicate that
interests of crowds of Web users are surprisingly predictable.

In summary, the models of attention dynamics considered in this paper provide simple yet reliable and theoretically well founded tools for Web trend analysis. They thus constitute new baselines for Web intelligence research that targets socio-economic questions. In particular, they provide baseline tools that help estimating the future success or customer adoption of particular services or Web-based businesses.

\section{Acknowledgments}

The work reported in this paper was carried out within the Fraunhofer / University of Southampton research project \emph{SoFWIReD} and funded by the Fraunhofer ICON initiative. Kristian Kersting was additionally supported by the Fraunhofer ATTRACT fellowship ``Statistical Relational Activity Mining''.

\bibliographystyle{abbrv}
\bibliography{literature}

\balancecolumns
\end{document}